\documentclass[fleqn,usenatbib]{mnras}

\usepackage{newtxtext,newtxmath}

\usepackage[T1]{fontenc}

\DeclareRobustCommand{\VAN}[3]{#2}
\let\VANthebibliography\thebibliography
\def\thebibliography{\DeclareRobustCommand{\VAN}[3]{##3}\VANthebibliography}

\usepackage{graphicx} 
\usepackage{amsmath}  
\usepackage{threeparttable} 
\usepackage{pdflscape} 

\usepackage{etoolbox, siunitx}
\newcommand{\cpd}{CPD-54 810}
\newcommand{\teff}{\ensuremath{T_\mathrm{eff}}}
\newcommand{\teffpri}{\ensuremath{T_\mathrm{eff,1}}}
\newcommand{\teffsec}{\ensuremath{T_\mathrm{eff,2}}}
\newcommand{\fbol}{\ensuremath{F_\mathrm{bol}}}

\newcommand{\tess}{\emph{TESS}}
\newcommand{\gaia}{\emph{Gaia}}

\newcommand{\lightkurve}{\textsc{lightkurve}}

\newcommand{\jktebop}{\textsc{jktebop}}

\DeclareSIUnit\Rsun{R_\odot}
\DeclareSIUnit\Msun{M_\odot} 
\DeclareSIUnit\Lsun{L_\odot} 
\DeclareSIUnit\gyr{Gyr}
\DeclareSIUnit\au{AU}

\defcitealias{2021MNRAS.500.4972R}{R21}
\defcitealias{2020MNRAS.497.2899M}{M20} 

\title[Fundamental Effective Temperature of CPD-54\,810]{Fundamental effective temperature measurements for eclipsing binary stars -- II. The detached F-type eclipsing binary CPD-54\,810}

\author[N. J. Miller et al.]{
N. J. Miller,$^{1}$\thanks{E-mail: n.j.miller1@keele.ac.uk}
P. F. L. Maxted,$^{1}$
D. Graczyk,$^{2}$
T. G. Tan$^{3}$
and J. Southworth$^{1}$
\\
$^{1}$Astrophysics Group, Keele University, Keele, Staffordshire, ST5 5BG, UK\\
$^{2}$Centrum Astronomiczne im. Miko\l{}aja Kopernika, Polish Academy of Sciences, Rabia\'{n}ska 8, 87-100, Toru\'{n}, Poland\\
$^{3}$Perth Exoplanet Survey Telescope, Perth, WA 6010, Australia
}

\date{Accepted YYYY MM DD. Received YYYY MM DD; in original form YYYY MM DD}

\pubyear{2022}

\begin{document}
\label{firstpage}
\pagerange{\pageref{firstpage}--\pageref{lastpage}}
\maketitle

\begin{abstract}
\cpd{} is a double-lined detached eclipsing binary containing two mid-F type dwarfs on an eccentric 26-day orbit.
We perform a combined analysis of the extensive photometry obtained by the \tess{} space mission along with previously published observations to obtain a full orbital and physical solution for the system. We measure the following model-independent masses and radii:
$M_1=1.3094\pm0.0051$\,\si{\Msun}, $M_2=1.0896\pm0.0034$\,\si{\Msun}, $R_1=1.9288\pm0.0030$\,\si{\Rsun}, and $R_2=1.1815\pm0.0037$\,\si{\Rsun}. 
We employ a Bayesian approach to obtain the bolometric flux for both stars from observed magnitudes, colours, and flux ratios. These bolometric fluxes combined with the stars' angular diameters (from $R_1$, $R_2$ and the parallax from \gaia{} EDR3) lead directly to the stars' effective temperatures: \teffpri{}$=6462\pm43$\,K, and \teffsec{}$=6331\pm43$\,K, with an additional systematic error of 0.8\% (13\,K) from the uncertainty in the zero-point of the flux scale. Our results are robust against the choice of model spectra and other details of the analysis.
\cpd{} is an ideal benchmark system that can be used to test stellar parameters measured by large spectroscopic surveys or derived from asteroseismology, and calibrate stellar models by providing robust constraints on the measured parameters. The methods presented here can be applied to many other detached eclipsing binary systems to build a catalogue of well--measured benchmark stars.

\end{abstract}

\begin{keywords}
stars: solar-type -- binaries: eclipsing -- stars: fundamental parameters
\end{keywords}



\section{Introduction}

Large uncertainties in the effective temperature (\teff{}) estimates for FGK stars are an obstacle to further progress in many areas of Galactic astrophysics. A recent review of industrial-scale measurements of stellar abundances by \citet{2019ARA&A..57..571J} found that most FGK-type stars, excluding the Sun, have derived effective temperatures accurate to 200$-$300\,K, with none more accurate than 50\,K. 
Uncertainties in \teff{} are the dominant source of uncertainty when calibrating stellar atmosphere models. A lack of reliable calibration stars thus undermines efforts in constraining several fundamental aspects of astrophysics, e.g. the estimation of stellar ages, which are needed to constrain models for planet \citep{2016A&A...587A..16V, 2018A&A...620A.168V} and galaxy formation \citep{2014ApJ...792..110V}.
Stellar spectroscopy, either performed by large surveys (e.g., RAVE \citep{2020AJ....160...83S}, SDSS \citep{2021arXiv211202026A}, LAMOST \citep{2012RAA....12..735D}, \gaia{}-ESO \citep{2012Msngr.147...25G}, APOGEE \citep{2017AJ....154...94M}, GALAH \citep{2018MNRAS.478.4513B}, 
the upcoming WEAVE \citep{2012SPIE.8446E..0PD} and 4MOST \citep{2019Msngr.175....3D}) or on an individual basis, needs a reliable \teff{} scale with which to calibrate stellar parameters. In these large surveys, data-driven approaches and machine learning methods are increasingly being used for the analysis. These are trained and calibrated on data with classical determinations of parameters. There is no physics in these data-driven methods so they must use benchmark calibration stars to establish how features in the data relate to astrophysical quantities, such as \teff{}.
There is therefore an urgent need for improved measurements of effective temperature for a large, representative sample of stars.

Substantial work has already been invested in calibrating the effective temperature scale for FGK-type main-sequence and subgiant stars. 
The testing and calibration of effective temperature estimates for these stars currently relies on measurements of angular diameter ($\theta$) for nearby stars using interferometry, and estimates of the bolometric flux (\fbol{}). For example, the \gaia{} FGK benchmark sample consists of 35 stars with \teff{} estimates derived using this approach \citep{2015A&A...582A..49H}, and is used widely by the community.
These stars are very bright, with V-band magnitude typically in the range 1-6 \citep{2014A&A...564A.133J}, which is significantly brighter than the magnitude limits for typical spectroscopic surveys. Since interferometric angular diameter measurements are limited to nearby stars with sufficiently large resolved discs, the sample of stars for which it is feasible to obtain a direct \teff{} estimate remains quite restricted. Consequently, there are gaps in the parameter space of \teff{} benchmark samples: most cool dwarfs and metal-poor stars are excluded.
An additional problem comes from the uncertainties present in angular diameter measurements, with repeated measurements of the same star showing variation larger than the quoted errors, often up to 5\,\%. For the \gaia{} benchmark sample, $\theta$ and \fbol{} are measured to 3\,\% and 5\,\% respectively, corresponding to uncertainties in \teff{} of 1.5\,\% and 1\,\%, i.e. approximately 100\,K for a solar-type star. \citet{2020arXiv201207957T} suggest that this uncertainty is even higher, with current $\theta$ and \fbol{} measurements carrying a systematic uncertainty floor in \teff{} of 2\,\%, corresponding to 120\,K for a solar-type star.
It is therefore important to pursue other ways to determine angular diameters to obtain robust \teff{} measurements to a higher accuracy and for a more representative sample of stars. 

Detached eclipsing binary stars (DEBs) are typically long-period systems in which the component stars can be assumed to have evolved separately. Apart from the Sun and a few nearby stars, DEBs are the only source of precise, model-independent mass and radius measurements for normal stars. 
With space-based photometry from \tess{} and spectroscopy from modern \'{e}chelle spectrographs it is possible to measure DEB component masses and radii to much better than 0.5\% \citep{2020MNRAS.498..332M} and thus surface gravity to better than 0.005 dex. 
DEBs with total eclipses are particularly useful as benchmark stars because we can obtain a `clean' spectrum of a single component star during totality, and use it to perform more detailed analyses of the abundances and spectroscopic parameters, and to perform spectral disentangling more easily.

The availability of \gaia{} parallaxes and high quality TESS light curves opens up a new method for obtaining angular diameters using DEBs. Combining precise measurements of the stellar radii with \gaia{} parallaxes makes it possible to measure the angular diameters for stars in DEBs to better than 1\%, and with a good knowledge of the bolometric flux \fbol{}, \teff{} to $\pm$50\,K or better.
In the first paper in this series \citep[][\citetalias{2020MNRAS.497.2899M} hereafter]{2020MNRAS.497.2899M} we demonstrated a new method for calculating \fbol{} for both stars in a DEB, which combines radius and parallax measurements with multi-band photometry of the eclipses to determine the overall shape of the flux distribution based on the observational data, rather than relying solely on model spectral energy distributions (SEDs). The method addresses the circular problem with SED fitting: the measurement of \teff{} relies on selecting an appropriate model, which in turn requires knowledge of \teff{}. In addition, it attempts to address the approximately 4\,\% difference in \fbol{} from integrating observed and model spectral energy distributions for K and M stars identified by \citet{2015A&A...582A..49H}.
In \citetalias{2020MNRAS.497.2899M}, we were able to measure fundamental effective temperatures for the well-studied DEB AI Phoenicis with very good precision: $6193 \pm 24$\,K for the F7\,V component and $5090 \pm 17$\,K for the K0\,IV component.

\cpd{} (also known as ASAS J051753-5406.0 or TYC 8511-888-1) is a moderately bright (V$=10.5$), totally-eclipsing detached eclipsing binary system first studied by \citet[][\citetalias{2021MNRAS.500.4972R} hereafter]{2021MNRAS.500.4972R}. They performed light curve and radial velocity fits to obtain masses and radii for both components, and obtained spectroscopic effective temperature estimates from disentangled optical spectra. Their analysis of the system suggests that the primary component is either an evolved main-sequence star or sub-giant of a late-F type ($M_1=1.311\pm0.035$\,\si{\Msun}, $R_1=1.935\pm0.020$\,\si{\Rsun}, \teffpri{}$=5980\pm205$\,K), while the secondary is a lower mass, early G-type main-sequence star ($M_2=1.093\pm0.029$\,\si{\Msun}, $R_2=1.181\pm0.014$\,\si{\Rsun}, \teffsec{}$=5850\pm190$\,K). These spectroscopic temperatures are $\sim$500\,K cooler than what we would expect from a preliminary look at the \gaia{} photometric colours. 

Here we present a re-analysis of all available observations of \cpd{}, to obtain new values for the masses and radii. Since the publication of \citetalias{2021MNRAS.500.4972R}, the number of \tess{} sectors containing observations of \cpd{} have more than doubled so we have been  able to improve the precision of the mass and radius measurements. In addition, we measure fundamental effective temperatures of both components, and draw conclusions about the evolutionary status of the system.
In contrast to AI Phoenicis, there are no published high-quality multi-band light curves of \cpd{} apart from \tess{}; it is thus more representative of the vast number of eclipsing binaries in the process of being identified by \tess{} and other large-scale photometric surveys. This makes it an interesting system to test whether the method introduced in \citetalias{2020MNRAS.497.2899M} will produce meaningful results for a large sample of poorly-studied DEBs and what other data are required to reach the levels of accuracy in mass, radius and effective temperature required for a particular system to be suitable as a benchmark system.

\section{OBSERVATIONS}

\subsection{\tess{} photometry}

The \tess{} satellite observed \cpd{} in the 2-minute cadence mode between 24 September 2018 and 17 July 2019, covering seven sectors (3-7, 10, 13). \tess{} returned to observe \cpd{} in the 10-minute cadence mode between 05 July 2020 and 13 January 2021, covering another five sectors (27, 30-33).

\subsection{Ground based photometry}

\cpd{} was observed by the All-Sky Automated Survey (ASAS) from 20 November 2000 to 12 October 2009 in the V band, with 698 good quality data points available in the ASAS Catalog of Variable Stars \citep[ACVS;][]{2002AcA....52..397P}. However, there were comparatively few good data points within the eclipses to obtain tight constraints on the orbital parameters or flux ratio of the binary. We used the times of primary minima present in the ASAS data to check the linear ephemeris of the system with a longer time baseline than otherwise possible. These were consistent with the other data but the lack of precision from the light curve fits led us to not include these data in our results.

We obtained photometric observations of \cpd{} with the Perth Exoplanet Survey Telescope (PEST) during the secondary eclipse on 04 January 2021. The PEST is a 0.3\,m telescope in Perth, Australia. At the time of these observations it was equipped with an SBIG ST-8XME camera and Astrodon B, V, Rc, Ic filters, giving an image scale of 1''.2 and a 31' x 21' field of view. Individual exposure times for the B, V, Rc, Ic bands were 120s, 60s, 30s, 60s respectively. Differential photometry was done with reference to an ensemble of comparison stars in the field. These photometric observations were reduced using the custom PEST pipeline\footnote{\url{http://pestobservatory.com/the-pest-pipeline/}}.
Magnitudes and errors for the PEST observations are given in Table \ref{tab:ground-based-photometry}.

In addition, \cpd{} was observed by the WASP-South instrument \citep{pollacco2006}, from 24 September 2012  to 19 November 2014, yielding a total of 24398 good quality data points.
For observations of \cpd{}, the WASP-South instrument was operated using Canon 85-mm f/1.2 lenses, 2k$\times$2k $e$2$V$ CCD detectors, and an r$^{\prime}$ filter. With these lenses the image scale was 33 arcsec/pixel. Fluxes were measured in an aperture with a radius of 132 arcsec and were processed with the SYSRem algorithm \citep{2005MNRAS.356.1466T} to remove instrumental effects.
Magnitudes and errors for WASP observations during and up to one day before and after eclipses are given in Table \ref{tab:ground-based-photometry}.

\begin{table}
    \centering
    \begin{tabular}{lccc}
    \hline
    \noalign{\smallskip}
    BJD & Filter & Magnitude & Error \\
    \noalign{\smallskip}
    \hline
    \noalign{\smallskip}
    2456195.4380815	&	r'	&	10.4497	&	0.0309	\\
    2456195.4384171	&	r'	&	10.4130	&	0.0292	\\
    2456195.4387528	&	r'	&	10.4207	&	0.0295	\\
    2456195.4427344	&	r'	&	10.4413	&	0.0288	\\
    2456195.4434172	&	r'	&	10.4338	&	0.0280	\\
    2456195.4474451	&	r'	&	10.4261	&	0.0270	\\
    2456195.4477808	&	r'	&	10.4150	&	0.0265	\\
    2456195.4481280	&	r'	&	10.4439	&	0.0264	\\
    2456195.4520517	&	r'	&	10.4251	&	0.0265	\\
    2456195.4523873	&	r'	&	10.4516	&	0.0264	\\
    \noalign{\smallskip}
    \hline
    \end{tabular}
    \caption{Ground-based photometric observations for \cpd{} from PEST (B, V, R, I) and WASP (r') observatories. For brevity, only WASP photometry of and around the eclipses used in the ephemeris calculation are provided. The full version of this table can be found in the online supplementary materials.}
    \label{tab:ground-based-photometry}
\end{table}

\begin{figure}
    \centering
    \includegraphics[width=0.475\textwidth]{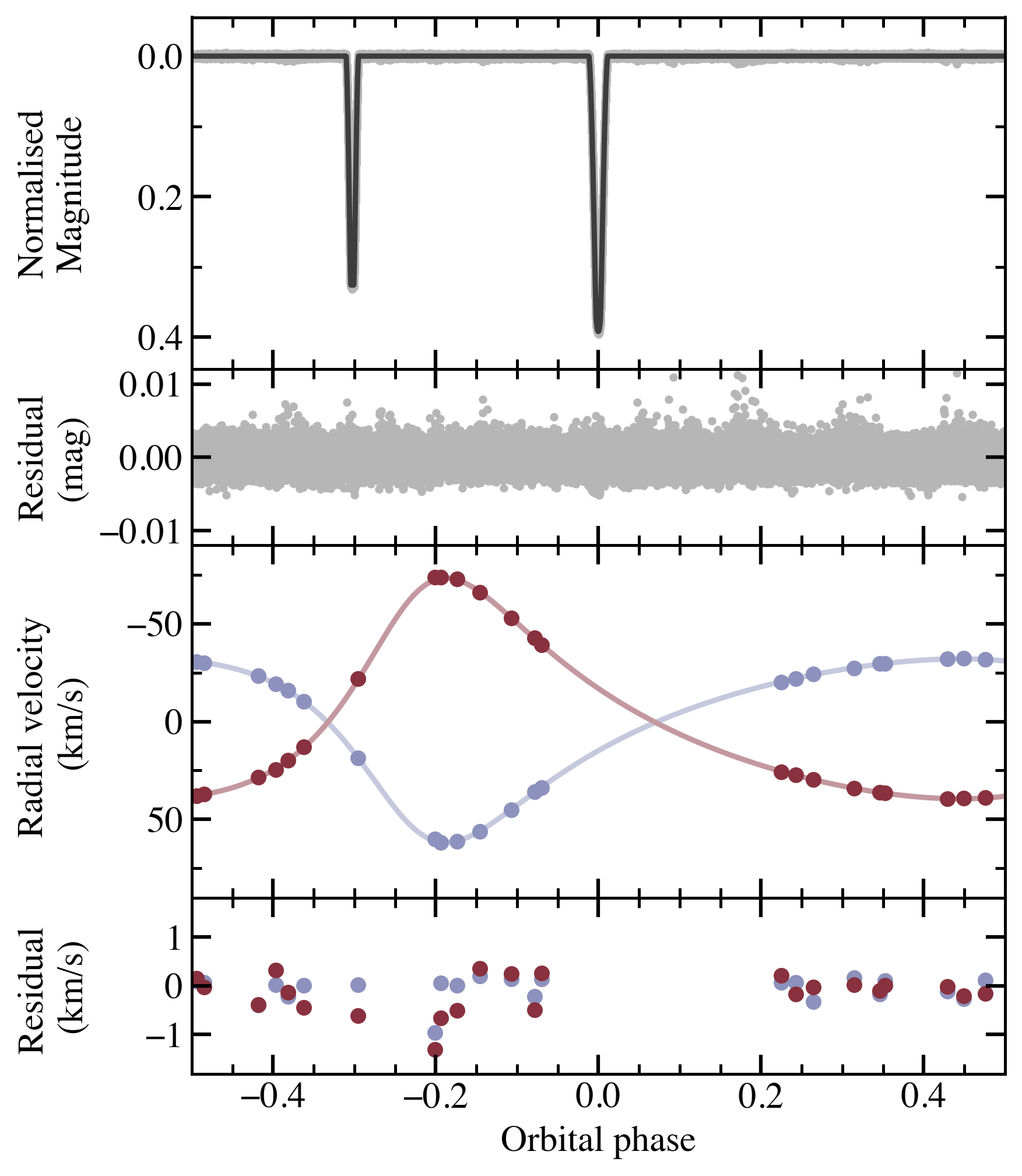}
    \caption{Top: \tess{} light curve consisting of all 2-minute cadence observations from Sectors 3-7, 10 and 13 with the \jktebop{} light curve solution and residual over the full orbital cycle. Bottom: Radial velocities extracted by \citetalias{2021MNRAS.500.4972R}, re-fitted using the radial velocity model in the \textsc{ellc} code, along with residuals from the best solution.} 
    \label{fig:rv-lc}
\end{figure}

\subsection{Catalog photometry} 

In our \teff{} analysis, we require magnitude measurements throughout the entire optical range, converted to the AB magnitude scale to allow us to compare observed magnitudes with synthetic magnitudes generated by the synthetic spectra (see Section \ref{sec:method-teffs}). For details on these transformations, see appendix A of \citet{2012PASP..124..140B}.

GALEX \citep{2005ApJ...619L...1M} observed \cpd{} in the FUV and NUV bands. Photometric response functions were taken from the GALEX web pages\footnote{\url{https://asd.gsfc.nasa.gov/archive/galex/tools/Resolution_Response/}} and the zero-point error from \citet{2014MNRAS.438.3111C}.
Our previous work on AI Phoenicis \citepalias{2020MNRAS.497.2899M} showed that more constraints on the ultraviolet and blue end of the optical range improve the reliability of the \teff{} measurements obtained. Unlike AI Phoenicis, \cpd{} does not have any archival space-based ultraviolet light curves and hence flux ratio. Therefore we chose to include u and v magnitudes from the SkyMapper survey \citep{2007PASA...24....1K}. We calculated a zero-point for each of these bands by calculating synthetic magnitudes for a set of CALSPEC \citep{2014PASP..126..711B} stars in the magnitude range 8\,--\,15 mag with SkyMapper u- and v- photometry.
We include G, BP and RP magnitudes and photometric zero points from \gaia{} Early Data Release 3 \citep[EDR3;][]{2021A&A...649A...1G} in our analysis. 
We included magnitudes in the J, H and Ks bands from 2MASS \citep{2006AJ....131.1163S}, with response functions obtained from the Explanatory Supplement to the 2MASS All Sky Data Release.\footnote{\url{https://old.ipac.caltech.edu/2mass/releases/allsky/doc/explsup.html}} The zero-points with their standard errors are taken from \citet{2018A&A...616L...7M}.
WISE magnitudes are taken from the All-Sky Release Catalog \citep{2012yCat.2311....0C}, with corrections to Vega magnitudes made as recommended by \citet{2011ApJ...735..112J}.

\subsection{Spectroscopic observations}\label{sec:obs-spec}

We used spectroscopic observations from the European Southern Observatory science archive facility\footnote{\url{http://archive.eso.org}} obtained with the 2.2-m MPG telescope equipped with the Fiberfed Extended Range Optical Spectrograph \citep[FEROS; R$\sim$48000;][]{1999Msngr..95....8K} to obtain measurements of the stars' metallicity and rotational velocity,
and the equivalent widths of the Na\,I\,D lines used to obtain an estimate of the interstellar reddening. \cpd{} was observed by FEROS on three occasions between 21 September 2012 and 25 September 2012 (Run ID: 089.D-0097(A), PI: He{\l}miniak) and three occasions between 28 December 2012 and 31 December 2012 (Run ID: 090.D-0061(A), PI: He{\l}miniak).

\subsection{Ground based radial velocity measurements}

We reanalysed the radial velocity measurements of \cpd{} extracted by \citet{2021MNRAS.500.4972R} to ensure that the parameters of the stars' spectroscopic orbits are consistent with the new results derived here from the light curves. The spectra from which these were measured were taken with FEROS (described in Section \ref{sec:obs-spec}), the Swiss 1.2-m Leonhard Euler Telescope with CORALIE \citep[R$\sim$60000;][]{2001Msngr.105....1Q} and the 1.5-m SMARTS telescope \citep{2010SPIE.7737E..1CS} at Cerro Tololo Inter-American Observatory with the CHIRON spectrograph in fiber mode (R$\sim$25000) and slit mode (R$\sim$90000).

\section{METHODS AND ANALYSIS}

\subsection{Reduction of photometric data}

The available \tess{} sectors were split into the 2-minute and 10-minute cadence modes. Sectors 4 and 32 were excluded from the analysis due to significant anomalous variation in the out-of-eclipse levels that appear to be due to instrumental effects. For the first set of \tess{} observations in the 2-minute cadence, we used target pixel files extracted by the Science Processing Operations Center (SPOC) pipeline \citep{2016SPIE.9913E..3EJ}. For the later set of sectors observed in the 10-minute cadence, we used target pixel files extracted from full frame image (FFI) files by the ``TESS-SPOC'' pipeline \citep{2020RNAAS...4..201C}. All \tess{} products were accessed from the Mikulski Archive
for Space Telescopes\footnote{\url{https://archive.stsci.edu/}} (MAST) via the \lightkurve{} package \citep{2018ascl.soft12013L}. Due to significant inconsistencies between the eclipse depths of the 2-minute and 10-minute cadence data sets for the pipeline-extracted light curves, we decided to perform our own simple aperture photometry. For this we used the pipeline-defined target aperture and a custom background aperture for each cadence, defined as the 20\% of pixels with the lowest flux. We removed systematics for each sector using cotrending basis vectors (CBVs), and applied corrections for crowding and fraction of flux in the aperture.
We then cleaned each sector by removing any data points with a poor quality flag, then normalised each sector by masking the eclipses, fitting a low-order polynomial to the out-of-eclipse continuum and dividing through the entire sector.

\subsection{Updated linear ephemeris}\label{sec:ephem}

We obtained a new measurement of the linear ephemeris using high quality observations of primary eclipses from WASP and \tess{}. For the \tess{} observations, we chose to only include eclipses for which the majority of the ingress and egress were observed. Despite some ASAS observations occurring further back in time than WASP observations, we chose not to include the ASAS light curves in our measurements due to its poor quality.
Times of mid eclipse were measured for each eclipse with \jktebop{}, fixing all other fit parameters to adopted values. From these times, given in Table \ref{tab:ephem} we measured the following linear ephemeris for the system:

$${\rm BJD}~T_{\rm mid} = 2458679.151318(12) + 26.13132764(11)\: E.$$

From performing the same analysis on secondary eclipses, we see no evidence of a third body or apsidal motion in the system. Fitting a quadratic ephemeris gives an upper limit on the rate of period change $|\dot{P}/P| < 5\times10^{-6}$.

\begin{table}
\caption{Times of mid eclipse for \cpd{}. The (O--C) residuals are from the linear ephemeris given in Section \ref{sec:ephem}. Details for the source of each eclipse used in the calculation are given, including the cadence of the \tess{} observations.}
\label{tab:ephem}
\begin{center}
    \begin{tabular}{@{}lrl}
        \hline
        \noalign{\smallskip}
        \multicolumn{1}{@{}l}{BJD--2450000} & (O--C) [s] & Source \\
        \noalign{\smallskip}
        \hline
        \noalign{\smallskip}
        6196.67563	$\pm$	0.00062	&$	37.4	$&	WASP	\\
        6954.48397	$\pm$	0.00092	&$	23.8	$&	WASP	\\
        8391.70667	$\pm$	0.00003	&$	-3.6	$&	TESS Sector 3	\\
        8443.96944	$\pm$	0.00004	&$	6.2	    $&	TESS Sector 5	\\
        8470.10068	$\pm$	0.00004	&$	-1.2	$&	TESS Sector 6	\\
        8496.23202	$\pm$	0.00004	&$	-0.8	$&	TESS Sector 7	\\
        8574.62604	$\pm$	0.00004	&$	2.6	    $&	TESS Sector 10	\\
        8679.15128	$\pm$	0.00004	&$	-3.1	$&	TESS Sector 13	\\
        9044.98993	$\pm$	0.00004	&$	1.8	    $&	TESS Sector 27	\\
        9123.38392	$\pm$	0.00004	&$	2.4	    $&	TESS Sector 30	\\
        9149.51519	$\pm$	0.00004	&$	-2.6	$&	TESS Sector 31	\\
        \noalign{\smallskip}
        \hline
    \end{tabular}
\end{center}
\end{table}

\subsection{Orbital and stellar parameters from TESS light curves}

We decided to re-analyse the light curves and radial velocities of \cpd{} using the \tess{} data that has become available since the analysis by \citetalias{2021MNRAS.500.4972R}. 
As is advised in \citet{2020MNRAS.498..332M}, it is good practice to carry out an independent analysis when performing light curve fitting at high precision. Therefore, we performed three independent analyses using different light curve fitting codes: \jktebop{} \citep{2013A&A...557A.119S}, {\tt ellc} \citep{2016A&A...591A.111M}, and the Wilson-Devinney (WD) code \citep{1971ApJ...166..605W}. A full description of the approach taken by each of the analyses is given in the Appendices, but we present a summary of the results of the light curve fits in Table \ref{compare-results}.

\begin{table}
\caption{Comparison of results for the analysis of the \tess{} light curves of \cpd{} using 3 different methods. Figures in parentheses give the standard error in the final digit of the preceding value.}
\label{compare-results}
\begin{center}
\begin{tabular}{@{}lrrrr}
\hline
\noalign{\smallskip}
\multicolumn{1}{@{}l}{Parameter} & \multicolumn{1}{c}{jktebop} & \multicolumn{1}{c}{WD} & \multicolumn{1}{c}{ellc} & \multicolumn{1}{c}{Adopted} \\
\noalign{\smallskip}
\hline
\noalign{\smallskip}
$r_1$      	       & 0.03893(2)     & 0.03886(7)	& 0.03890(1)  & 0.03891(4)\\
$r_2$      	       & 0.02379(4)     & 0.02395(10)	& 0.02384(2)  & 0.02383(7)\\
$i$ [$^{\circ}$]   & 89.72(2)       & 89.83(3)	    & 89.742(9)   & 89.76(5) \\
e                  & 0.3686(1)      & 0.3691(1)	    & 0.36859(6)  & 0.3688(4)\\
$\omega$ [$^{\circ}$] & 327.02(3)  & 327.01(2) & 326.83(3) & 329.96 (9) \\ 
$L_2/L_1$  	       & 0.350(2)       & 0.3596(42)    & 0.3535(9)  & 0.3534(44) \\
$\ell_3 $  	       & 0.002(4)       & 0.013(3)      & 0.009(2)   & 0.008(5) \\
\noalign{\smallskip}
\hline
\end{tabular}
\end{center}
\end{table}

 The adopted values in Table~\ref{compare-results} were calculated using the weighted mean and weighted sample standard deviation assuming that each of the three values from the \jktebop{}, WD and {\tt ellc} analyses are affected by the same systematic error $\sigma_{\rm sys}$ added in quadrature to the standard errors quoted in the three input values. The value of $\sigma_{\rm sys}$ was adjusted such that the weighted mean value as a model for the three input values has a reduced chi-square value $\chi^2_r = 1$. 

Masses and radii were then calculated in nominal solar units \citep{2016AJ....152...41P} using the Newtonian gravitational constant value $G=6.67408\times10^{-11} {\rm m}^3\,{\rm kg}^{-1}\,{\rm s}^{-2}$ (CODATA 2014 value). The adopted values of $K_1$ and $K_2$ used in this calculation were taken from the simple Keplerian orbital solution described in Appendix \ref{jktebop}. A Monte Carlo method was used for the calculation of the standard errors. The masses and radii for the stars in \cpd{} are given in Table~\ref{FundamentalParams}. For comparison, we also list in Table~\ref{FundamentalParams} the absolute parameters from \citetalias{2021MNRAS.500.4972R}. The precision of the mass measurements has been substantially improved, even though these are mostly determined by the same radial velocity measurements as \citetalias{2021MNRAS.500.4972R}. We were not able to reproduce the large error on the eccentricity obtained by \citetalias{2021MNRAS.500.4972R} ($e=0.367\pm 0.020$). This is the dominant contributor to the errors on the masses obtained by that study. 


\subsection{Flux ratios from TESS and PEST light curves}

\begin{figure}
    \centering
    \includegraphics[width=0.475\textwidth]{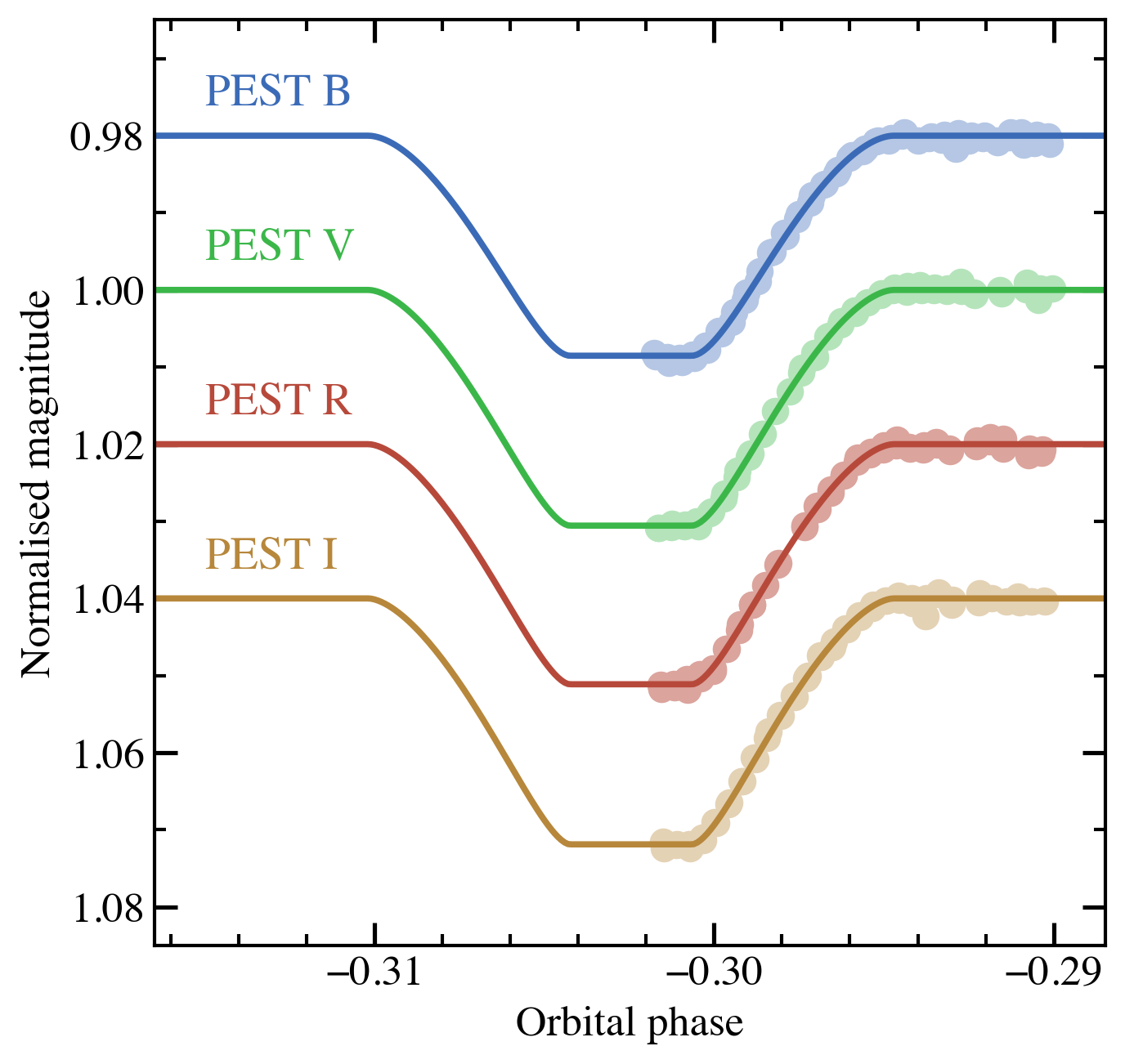}
    \caption{Photometric observations taken by the PEST observatory in BVRI bands along with the best \jktebop{} fit.}
    \label{fig:pest}
\end{figure}

The \tess{} flux ratios used in the \teff{} analysis were taken from the adopted light curve fit using \jktebop{}, as described in Section \ref{jktebop}. The adopted value and error, from the standard deviation of the eight subsets, are given in Table \ref{DataTable}.
The PEST flux ratios were calculated by fitting each light curve in \jktebop{}. Due to the limited phase coverage of the observations, we only allowed the surface brightness ratio $J$ and light scale factor to vary. We fixed the quadratic limb darkening coefficients for each filter to those described in \citet{2000A&A...363.1081C} and fixed all other parameters to the adopted values from the \tess{} fits. We used the MC methods in \jktebop{} to perform a fit for each light curve over 1000 simulations perturbing each observation randomly by its standard error to estimate the uncertainty on the flux ratio due to noise in the light curve. We also performed fits to the light curve with  each of the parameters $r_{\rm sum}$, $k$,  $i$, $\ell_3$ perturbed by their standard errors in order to quantify the uncertainty on the flux ratio due to errors on these parameters. Similarly, we also performed fits with the linear limb darkening coefficients for each star perturbed by a somewhat arbitrary error estimate of 0.1. The best values for the flux ratio (calculated from the surface brightness ratio) are given in Table \ref{DataTable} with all the contributions to the uncertainty added in quadrature. The light curve fits can be seen in Figure \ref{fig:pest}.

\subsection{Estimate of the interstellar extinction}\label{sec:extinction}

\begin{figure}
    \centering
    \includegraphics[width=0.475\textwidth]{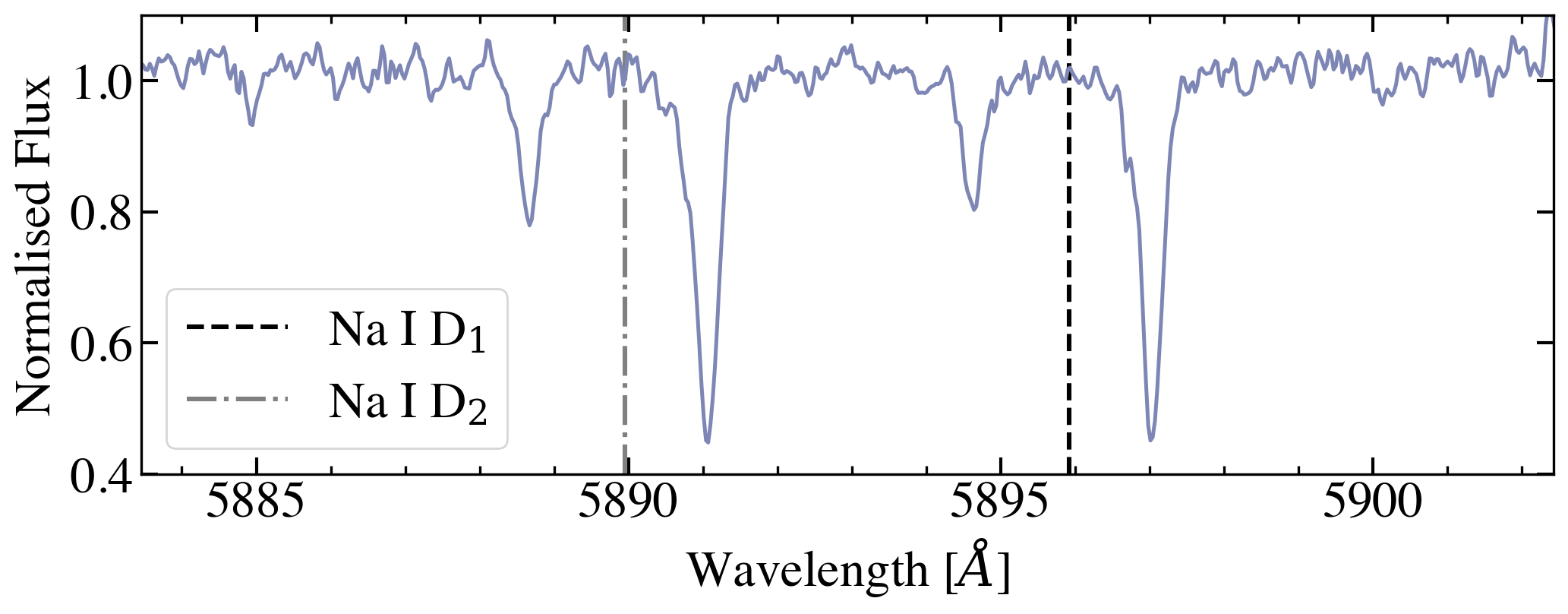}
    \caption{The FEROS spectrum with the largest wavelength spacing between stellar lines about the rest wavelength of the Na I D1 and D2 lines. The rest wavelengths of the two lines are marked with vertical lines.} 
    \label{fig:feros}
\end{figure}

An accurate measurement of a star's effective temperature based on photometry requires a robust estimate of the interstellar reddening. The importance of having a suitable prior on the interstellar reddening was demonstrated in Paper I \citepalias{2020MNRAS.497.2899M}, where placing no prior increased the uncertainty on the derived effective temperatures by 400\%. 
Relationships have previously been established between the equivalent widths of selected interstellar absorption lines such as the Na~I doublet and K~I line. The empirical relations established in \citet{2012MNRAS.426.1465P} are calibrated using spectra of quasars and galaxies, and hence are not well-constrained in the regime of E(B$-$V) < 0.01, which is where we would expect our relatively local ($\sim$380\,pc) system to lie. However, the approach taken by \citet{1997A&A...318..269M} uses a sample of O- and early B-type stars with a range of E(B$-$V) values from $0.0-1.6$ \citep{1994BaltA...3..158S}. In general, equivalent width $W$ of an interstellar absorption line is related to E(B$-$V) by 
$$W=\alpha \sum_{n=1}^{\infty}(-1)^{n-1} \frac{\left(\beta E_{B-V}\right)^{n}}{n ! \sqrt{n}},$$
where the constants $\alpha=0.354\pm 0.01$\,\AA\ and $\beta=11.0\pm 1.0$ for the Na I D1 line.
FEROS spectra of CPD-54\,810 were obtained from the ESO Science Archive Facility and the regions around the Na I doublet are shown in Figure \ref{fig:feros}. 
We fitted the Na I D1 line in each spectrum with grids of Gaussian models centred on the rest wavelength and spanning the $\pm 25$\,km/s velocity range of local clouds in the ISM \citep{2011ARA&A..49..237F}. Taking a mean equivalent width of $0.0074\pm0.0006$\,\AA
, we used a Bayesian approach to find a best fit and hence E(B$-$V) estimate for the system, exploring the posterior distribution of the model with MCMC methods to obtain a robust error. For \cpd{} we obtain a reddening estimate of E(B$-$V) = $0.002 \pm 0.012$, which includes an additional error of 0.011 from the scatter of the \citet{1997A&A...318..269M} relation.
Other measurements of reddening are available for \cpd{}, for example, StarHorse \citep{2022A&A...658A..91A} provides $A_V=0.236\pm0.140$, and \citet{2020AJ....159...84B} provides E(BP$-$RP) $= 0.0087\pm0.0692$. These estimates are typically slightly larger than the value measured from the Na~I~D1 line, but also carry a larger uncertainty. We choose to use the empirical measurement of E(B$-$V) as a prior in our analysis of \cpd{} since the depth of the interstellar absorption lines in the FEROS spectrum do not suggest a significantly higher value.

\subsection{Metallicity estimate}

It was necessary to obtain an estimate of the metallicity [Fe/H] for \cpd{} in order to select a reasonable model SED to use in the \teff{} calculation. 
We generated a grid of synthetic spectra for both stars over the metallicity range [Fe/H] = ($-0.6$, $-0.4$, $-0.2$, 0.0, 0.2) using the \textsc{ispec} \citep{2014A&A...569A.111B,2019MNRAS.486.2075B} implementation of the \textsc{turbospectrum} code \citep{2012ascl.soft05004P}. We used the MARCS grid \citep{2008A&A...486..951G} and solar abundances from \citet{2007SSRv..130..105G}. We fixed the surface gravity to the values given in Table~\ref{FundamentalParams} and fixed the effective temperatures to the values 6500\,K and 6350\,K, consistent with the values we derive below. Following \citet{2005ApJS..159..141V} we assumed a value of $v_{\rm mic}=0.85$\,km/s for both stars, and used their Equation (1) to obtain estimates for the macroturbulence velocities from our \teff{} estimates: $v_{\rm mac, 1}=2.84$\,km/s, $v_{\rm mac, 2}=3.05$\,km/s. 
We synthesised a grid of synthetic combined spectra, by shifting by radial velocities and co-adding the primary and secondary spectra. This allowed us to directly compare the observed FEROS spectra and synthetic spectra.
We iterated over the list of unblended Fe I and Fe II lines presented in \citet{2017MNRAS.469.4850D} and noted which of the synthetic grid best matched the depth of the Fe line. Any lines that were not present or blended were not included in the analysis. We took an average of the measured metallicities and obtained an estimate of ${\rm[Fe/H]} = 0.0 \pm 0.2$.

\begin{figure}
    \centering
    \includegraphics[width=0.475\textwidth]{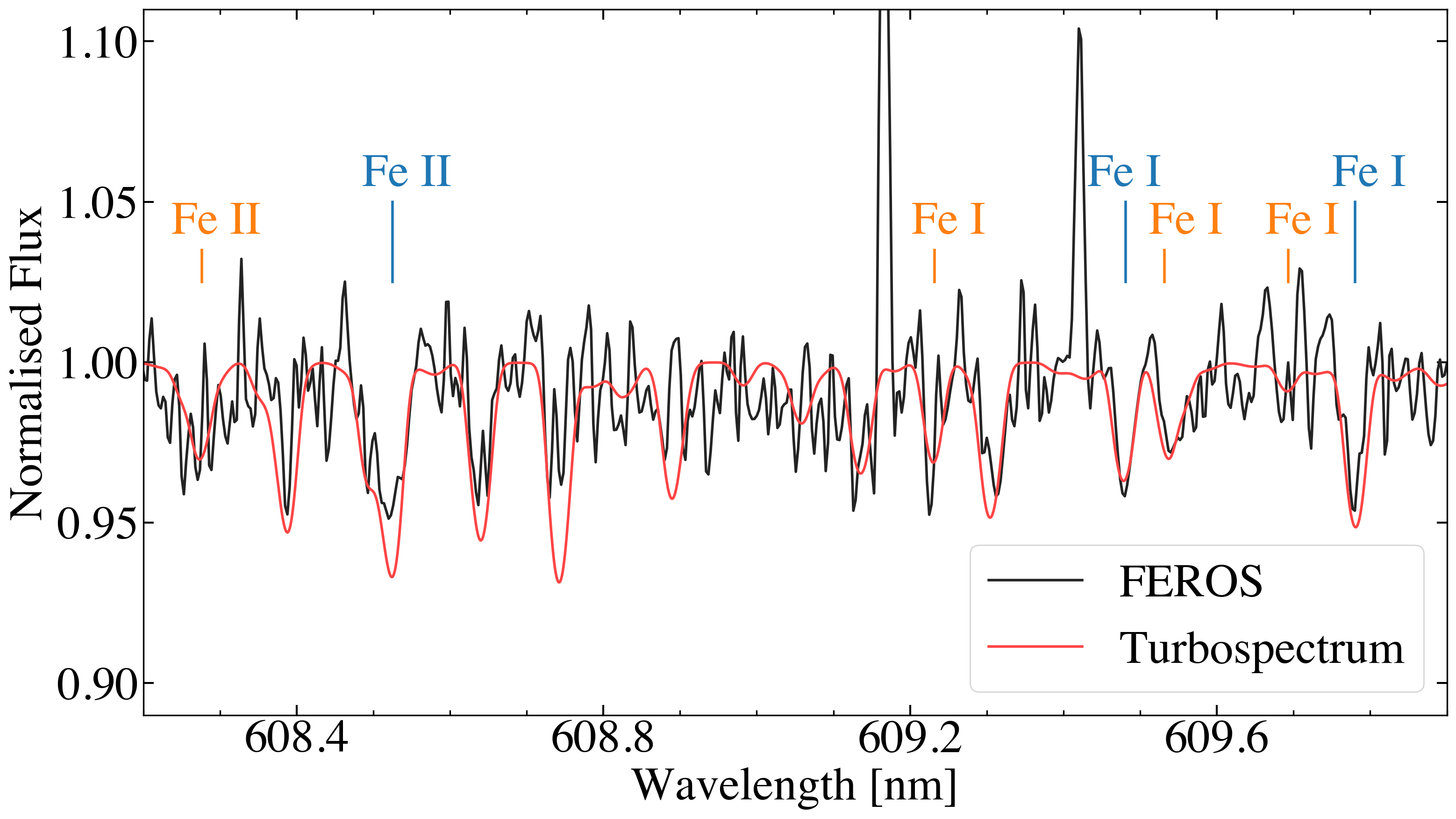}
    \caption{Observed FEROS spectrum of \cpd{} compared to the synthetic \textsc{turbospectrum} generated from the best \teff{}, $\log g$ and [M/H] estimates from our analysis. Prominent iron lines from \citet{2017MNRAS.469.4850D} used to estimate the metallicity of the system are noted above the continuum, shifted to the velocities of the primary (blue) and secondary (orange) components. }
    \label{fig:synthetic_spectrum}
\end{figure}

\subsection{Effective temperatures}\label{sec:method-teffs}

\begin{table}
\caption{Observational data for \cpd{} used in our \teff{} analysis.}
\label{DataTable}
\begin{center}
    \begin{tabular}{@{}lrl}
        \hline
        \noalign{\smallskip}
        \multicolumn{1}{@{}l}{Quantity}&
        \multicolumn{1}{l}{\hspace{6pt}Value}&
        \multicolumn{1}{l}{Source}\\
        \noalign{\smallskip}
        \hline
        \noalign{\smallskip}
        Parallax, $\varpi$ & $2.631 \pm 0.020$\,mas & \gaia{} EDR3$^1$ \\
        \noalign{\smallskip}
        Radius, R$_1$ & $1.9288 \pm 0.0030$\,R$_{\odot}$ & This work$^2$\\
        Radius, R$_2$ & $1.1815 \pm 0.0037$\,R$_{\odot}$ &~~~~" \\
        \noalign{\smallskip}
        \hline
        \noalign{\smallskip}
        \multicolumn{3}{@{}l}{Apparent magnitude} \\ 
        ~FUV  &$19.527 \pm 0.206$& GALEX  \\
        ~NUV  &$14.187 \pm 0.009$& ~~~"  \\
        ~u    &$11.770 \pm 0.009$& SkyMapper \\
        ~v    &$11.312 \pm 0.011$& ~~~" \\
        ~G    &$10.343 \pm 0.003$& \gaia{} EDR3 \\
        ~BP   &$10.576 \pm 0.003$& ~~~" \\
        ~RP   &$9.958 \pm 0.004$& ~~~" \\
        ~J    &$9.554 \pm 0.027$& 2MASS  \\
        ~H    &$9.380 \pm 0.026$& ~~~"  \\ 
        ~Ks   &$9.283 \pm 0.023$& ~~~"  \\
        ~W1   &$9.218 \pm 0.023$& WISE  \\ 
        ~W2   &$9.240 \pm 0.020$& ~~~"  \\
        ~W3   &$9.241 \pm 0.021$& ~~~"  \\
        ~W4   &$8.973 \pm 0.254$& ~~~" \\
        \noalign{\smallskip}
        \hline
        \noalign{\smallskip}
        \multicolumn{3}{@{}l}{Flux ratios} \\
        ~B    &$0.3299 \pm 0.0012$& This work  \\
        ~V    &$0.3413 \pm 0.0008$& ~~~"  \\ 
        ~R    &$0.3475 \pm 0.0009$& ~~~"  \\
        ~I    &$0.3517 \pm 0.0008$& ~~~"  \\
        ~TESS &$0.3517 \pm 0.0009$& ~~~"\\ 
        \noalign{\smallskip}
        \hline
        \noalign{\smallskip}
        \multicolumn{3}{@{}l}{Derived quantities} \\
        $\theta_1$ & $0.04720 \pm 0.00040$\,mas  & \protect{$2R_1 \varpi$} \\
        $\theta_2$ & $0.02891 \pm 0.00024$\,mas  & \protect{$2R_2 \varpi$} \\
        \noalign{\smallskip}
        \hline
    \end{tabular}
\begin{tablenotes}
$^1$Including correction from \citet{2022MNRAS.509.4276F}.\\
$^2$Including correction from apparent disc radius to Rosseland radius.
\end{tablenotes}
\end{center}
\end{table}

We use an approach based on the Stefan-Boltzmann law to obtain independent, fundamental effective temperatures for both components of an eclipsing binary system.
For a detached, non-interacting binary star system at distance $d$, i.e. with parallax $\varpi=1/d$, where each star has angular diameter $\theta = 2 R\varpi$, the total flux of the binary corrected for extinction at the top of the Earth's atmosphere is 
$$f_{0,b}= f_{0,1}+f_{0,2}=\frac{\sigma_{\rm SB}}{4}\left[\theta_1^2{\rm T}_{\rm eff,1}^4 + \theta_2^2{\rm T}_{\rm eff,2}^4\right],$$
where $\sigma_{\rm SB}$ is the Stefan-Boltzmann constant.
The radius $R$ used in the calculation of $\theta$ is the Rosseland radius, which is obtained by applying a correction to the photometric radius by noting the difference between the two radii for the Sun found by \citet{2009EGUGA..11.3961H} and scaling it to the appropriate \teff{} for the stars in \cpd{} using values measured by \citet{2017AJ....154..111M}. This correction is on the order of the atmospheric scale height, so is only significant for stars with very precise radii measured from the light curves.
The parallax for \cpd{} is taken from \gaia{} EDR3 with zero-point corrections from \citet{2022MNRAS.509.4276F}.
All of these quantities are known or can be measured for \cpd{} if we can independently and accurately obtain the integrated fluxes $f_{0,1}$, $f_{0,2}$ for both stars. This can be done by using observations of apparent magnitudes at ultraviolet, visible and infrared wavelengths, and since the light curve of \cpd{} shows total eclipses, it is possible to obtain a reliable estimate of the flux ratio in several photometric bands. 
To obtain reliable integrated fluxes for \cpd{}, we use the method first described in Paper I \citepalias{2020MNRAS.497.2899M}, which aims to avoid the caveats of simple SED fitting by balancing the observational constraints from photometry with the small-scale spectral features provided by the model SED. The method uses Legendre polynomials to distort the model SEDs based on the photometry, such that the large-scale shape of the flux integrating functions are determined by the data rather than the choice of model. 
The method uses \textsc{emcee} to sample the posterior distribution of $P(M|D)\propto P(D|M)P(M)$ for the model parameters $M$ with prior $P(M)$ given the data, $D$ (observed magnitudes and flux ratios).
The model parameters are $$M = \left({\rm T}_{\rm eff,1},  {\rm T}_{\rm eff,2}, \theta_1, \theta_2, {\rm E}({\rm B-V}), \sigma_{\rm ext}, \sigma_{\ell},  d_{1,1}, \dots, d_{2,1}, \dots\right).$$
The prior $P(M)$ is calculated as a combination of the priors on the near-infrared flux ratios (Section \ref{sec:NIR-FRPs}), ratio of the stellar radii (Section \ref{sec:k-ratio}) and a Gaussian prior on the interstellar extinction. The hyper-parameters $\sigma_{\rm ext}$ and $\sigma_{\ell}$ take into account additional uncertainties in the synthetic magnitudes and flux ratios, respectively, due to errors in the zero-points and response functions of the photometric passbands, errors in the SED models, or stellar variability.
The distortion function $\Delta_i$ for each star applied to a given model SED (to calculate synthetic photometry for a given \teff{}) is a linear superposition of Legendre polynomials in wavelength with coefficients for star 1 $d_{1,1}, d_{1,2}, \dots$, and similarly for star 2. The number of coefficients $N_{\Delta}$ can be varied, such that the optimal number can be found. The distorted model SED for each star is then normalised and can be integrated to calculate the total bolometric flux and synthetic photometry for each star. The effective temperatures derived using this method are based on the angular diameter and integrated stellar flux calculated using distortion to include the realistic stellar absorption features from the models but to allow the overall shape to be determined by the observed magnitudes, and thus much of the dependence on models that SED fitting suffers from is alleviated.

\subsubsection{Priors on infrared flux ratios}\label{sec:NIR-FRPs}
We do not have any direct measurements of the binary flux ratio at wavelengths longer than 1\,$\mu$m for \cpd{}. If there is no constraint placed on the flux for both stars in the near-infrared (NIR), the distortion functions could allow for models where the flux is unrealistically high or low. 
Following from Paper I \citepalias{2020MNRAS.497.2899M}, we note that for solar-type stars, there is a well-defined relationship between \teff{} and the NIR flux compared to total optical flux that shows little dependence on $\log g$ or [M/H].
Therefore, assuming that the stars in \cpd{} behave like other dwarf and subgiant FGK-type stars in the solar neighbourhood, we can put some constraints on the flux ratio in the 2MASS J, H, Ks and WISE W1, W2, W3, W4 bands. 
Using stars from the Geneva-Copenhagen survey \citep{2009A&A...501..941H,2011A&A...530A.138C} that are present in both 2MASS \citep{2006AJ....131.1163S} and WISE \citep{2012yCat.2311....0C} catalogs, we defined relations between \teff{} and (V--X) colours for each NIR bandpass.
We defined separate relations for the two stars, based on two subsets of stars with similar properties to each component of \cpd{}. We restricted both subsets to an interstellar reddening range of $\rm{E(B-V)}<0.01$, with the primary sample further restricted to $5800<{\rm\teff{}}<6800$\,K and $3.5<\log g<4.5$, and the secondary sample restricted to $5500<{\rm\teff{}}<6600$\,K and $3.8<\log g<4.8$. These relations are given in Table \ref{tab:nir-frps}.

\begin{table*}
\caption{Quadratic colour--\teff{} relations used to place Gaussian priors on the near-infrared flux ratio for \cpd{}, and the uncertainty on the colour for each.}
\label{tab:nir-frps}
\begin{center}
    \begin{tabular}{@{}lllll}
        \hline
        \noalign{\smallskip}
        \multicolumn{1}{@{}l}{Colour} & Primary & Error & Secondary & Error \\
        \noalign{\smallskip}
        \hline
        \noalign{\smallskip}
        V$-$J	&	$	0.000	X_1^2	-0.417	X_1+	0.965	$	&	0.042	&	$	0.000	X_2^2	-0.435	X_2+	1.073	$	&	0.042	\\
        \noalign{\smallskip}
        V$-$H	&	$	0.050	X_1^2	-0.555	X_1+	1.173	$	&	0.044	&	$	0.064	X_2^2	-0.585	X_2+	1.315	$	&	0.044	\\
        \noalign{\smallskip}
        V$-$Ks	&	$	0.066	X_1^2	-0.576	X_1+	1.238	$	&	0.039	&	$	0.085	X_2^2	-0.616	X_2+	1.387	$	&	0.039	\\
        \noalign{\smallskip}
        V$-$W1	&	$	0.046	X_1^2	-0.582	X_1+	1.286	$	&	0.061	&	$	0.095	X_2^2	-0.623	X_2+	1.434	$	&	0.061	\\
        \noalign{\smallskip}
        V$-$W2	&	$	0.060	X_1^2	-0.576	X_1+	1.277	$	&	0.097	&	$	0.050	X_2^2	-0.599	X_2+	1.424	$	&	0.097	\\
        \noalign{\smallskip}
        V$-$W3	&	$	0.074	X_1^2	-0.575	X_1+	1.235	$	&	0.053	&	$	0.112	X_2^2	-0.624	X_2+	1.383	$	&	0.053	\\
        \noalign{\smallskip}
        V$-$W4	&	$	0.098	X_1^2	-0.560	X_1+	1.281	$	&	0.092	&	$	0.106	X_2^2	-0.613	X_2+	1.426	$	&	0.092	\\
        \noalign{\smallskip}
        \hline
    \end{tabular}
\end{center}
\end{table*}

\subsubsection{Priors on ratio of the stellar radii}\label{sec:k-ratio}
\cpd{} is a totally-eclipsing system, which means that we have a very good estimate of the ratio of the fractional stellar radii $k$ from the \tess{} light curves. We apply an additional prior to the \teff{} fitting method to constrain the parameter space to a realistic solution.

\subsubsection{Application of the method to \cpd{}}
\begin{figure}
    \centering
    \includegraphics[width=0.475\textwidth]{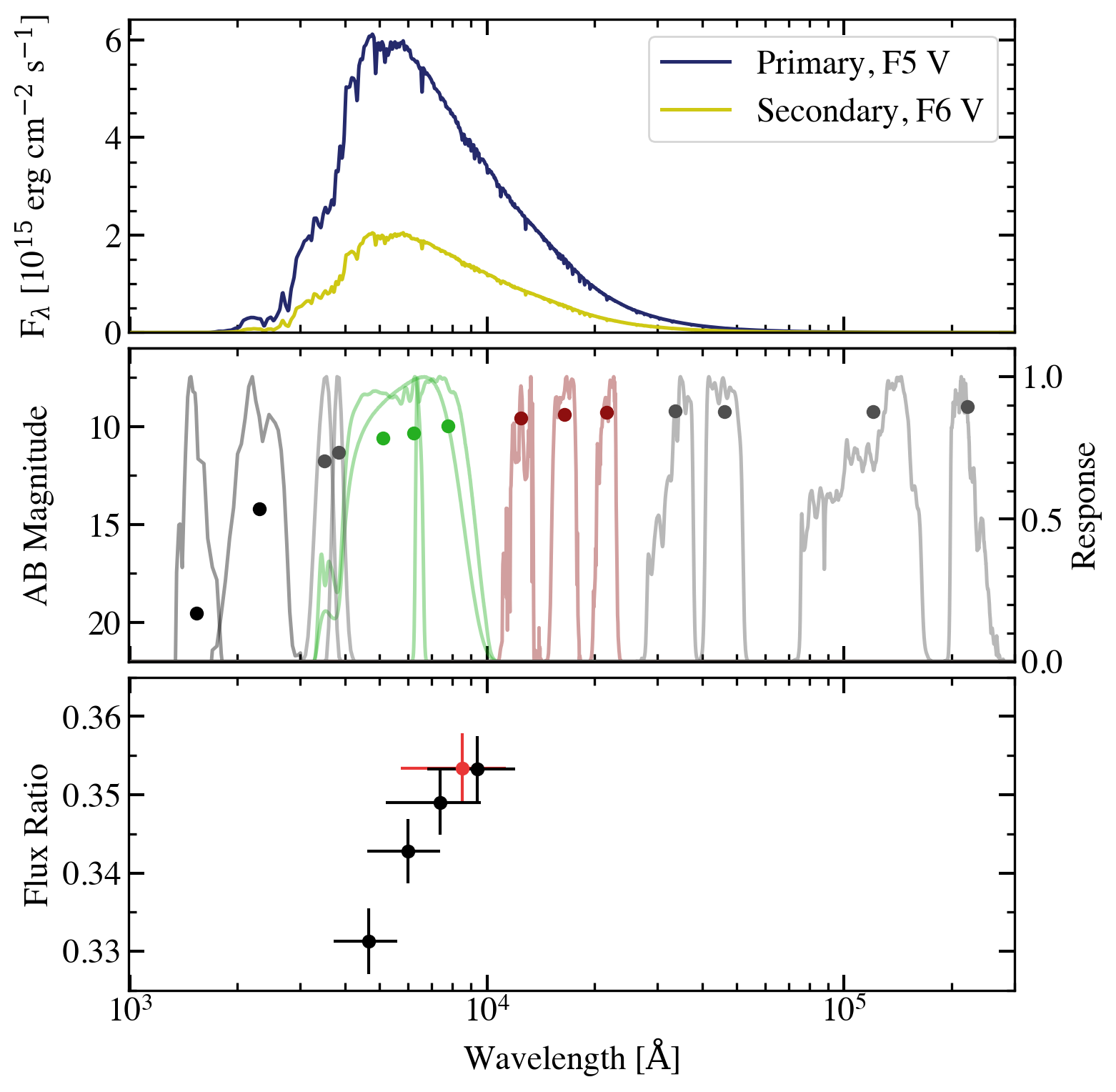}
    \caption{Summary of the photometric information used to derive \teff{}. Top: BT-Settl spectral energy distributions for the two component stars of \cpd{} with solar abundances from \citet{2009ARA&A..47..481A}, where \teffpri{}$=6450$\,K, $\log g_1=3.98$\,dex, [Fe/H]$_1=0.0$, and \teffsec{}$=6300$\,K, $\log g_2=4.33$\,dex. [Fe/H]$_2=0.0$. Each SED is scaled by the fractional radii of the two stars. 
    Middle: Observed AB magnitudes used to constrain the overall shape of the bolometric flux integrating functions in the \teff{} fit, and their photometric response functions. 
    Lower: Flux ratios obtained from light curve fits of the PEST (black) and \tess{} (red) data, where the x-error is the wavelength span of the filter bandpass.}
    \label{fig:input_photometry}
\end{figure}

For \cpd{}, we use BT-Settl model atmospheres \citep{2013MSAIS..24..128A} accessed via the Spanish Virtual Observatory\footnote{\url{http://svo2.cab.inta-csic.es/theory/newov2/index.php?models=bt-settl}} 
to calculate SEDs for both stars, using linear interpolation to obtain a model for each star with the parameters: $T_{\rm mod,1}=6450$\,K, $\log g_{\rm mod, 1}=3.98$, $T_{\rm mod,2}=6300$\,K, $\log g_{\rm mod, 2}=4.33$, and the same composition $[{\rm Fe/H}] = 0.0$, $[{\rm \alpha/Fe}] = 0.0$ for both components. The model SEDs, along with the observed magnitudes and flux ratios used in the analysis of \cpd{}, are shown in Figure \ref{fig:input_photometry}. The predicted apparent magnitudes and flux ratios along with their photometric zero-point errors for our adopted values of \teff{} fit are given in Table \ref{Residuals}, and are compared with the observed photometry. 

\begin{figure*}
    \centering
    \includegraphics[width=0.99\textwidth]{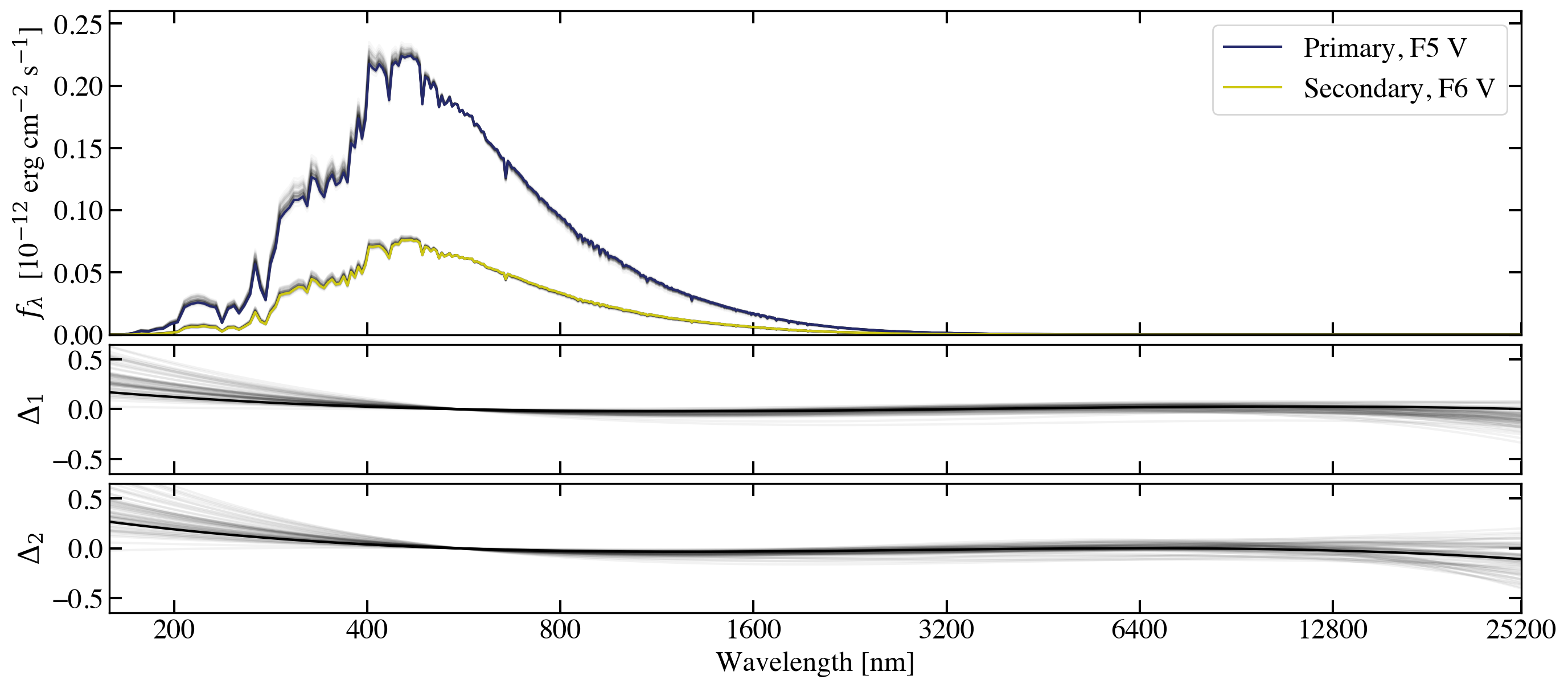}
    \caption{Integrating functions and distortion polynomials for our adopted \teff{} solution. Top: Best log-likelihood integrating functions of the two stars used to obtain the best values for \teffpri{} and \teffsec{}. Middle: The distortion functions applied BT-Settl input model SEDs for the primary star, showing best log-likelihood fit (dark line) and all other solutions (faint grey lines). Lower: Same, but for the secondary star.} 
    \label{fig:distort-plot}
\end{figure*}

We ran 16 different versions of the \teff{} analysis with 256 walkers over 1000 steps, with a burn-in of 1000 steps, to experiment with different input models, different numbers of distortion coefficients, and removing priors and observational data. Convergence of the fits were checked by a visual inspection of the trail plots. The details of each of these are discussed in detail in Section \ref{sec:teff-discussion}. The spectral energy distribution for the adopted fit is shown in Figure \ref{fig:distort-plot} and our best estimates for the stars' effective temperatures are given in Table \ref{FundamentalParams}.
The errors quoted in Table \ref{RunTable} do not account for the systematic error present due to uncertainties in the calibration of the CALSPEC flux scale \citep{2014PASP..126..711B}. For \cpd{}, this error is an additional 13\,K for both components.

\begin{table}
\caption{Predicted data values and residuals for the best-fit model from Run A. The predicted apparent magnitudes are quoted together with the error on the zero-point.}
\label{Residuals}
\begin{center}
    \begin{tabular}{@{}lrr}
        \hline
        \noalign{\smallskip}
        \multicolumn{1}{@{}l}{Parameter} & \multicolumn{1}{l}{\hspace{6pt}Value} & \multicolumn{1}{l}{\hspace{6pt}Residual} \\
        \noalign{\smallskip}
        \hline
        \noalign{\smallskip}
        \multicolumn{3}{@{}l}{Apparent magnitude} \\ 
        FUV     &$19.762\pm 0.374$&$ -0.235\pm 0.427$\\
        NUV     &$14.142\pm 0.883$&$ +0.045\pm 0.883$\\
        u       &$12.288\pm 0.219$&$ -0.518\pm 0.219$\\
        v       &$11.442\pm 0.102$&$ -0.130\pm 0.103$\\
        G       &$10.340\pm 0.003$&$ +0.003\pm 0.004$\\
        BP      &$10.573\pm 0.003$&$ +0.003\pm 0.004$\\
        RP      &$9.960\pm 0.004$&$ -0.003\pm 0.005$\\
        J       &$9.533\pm 0.005$&$ +0.021\pm 0.027$\\
        H       &$9.339\pm 0.005$&$ +0.041\pm 0.026$\\
        Ks      &$9.283\pm 0.005$&$ +0.000\pm 0.024$\\
        W1      &$9.245\pm 0.002$&$ -0.027\pm 0.023$\\
        W2      &$9.239\pm 0.002$&$ +0.001\pm 0.020$\\
        W3      &$9.224\pm 0.002$&$ +0.017\pm 0.021$\\
        W4      &$9.284\pm 0.002$&$ -0.311\pm 0.254$\\
        \noalign{\smallskip}
        \hline
        \noalign{\smallskip}
        \multicolumn{3}{@{}l}{Observed flux ratios} \\ 
        B     &0.332  &$-0.001 \pm 0.004$\\
        V     &0.343  &$+0.000 \pm 0.004$\\
        R     &0.351  &$-0.002 \pm 0.004$\\
        I     &0.355  &$-0.002 \pm 0.004$\\
        \tess{}  &0.353  &$+0.000 \pm 0.004$\\
        \noalign{\smallskip}
        \hline
        \noalign{\smallskip}
        \multicolumn{3}{@{}l}{Predicted flux ratios} \\ 
        J & 0.361 &$ -0.001\pm0.020$\\
        H & 0.369 &$ -0.004\pm0.021$\\
        Ks& 0.369 &$ -0.003\pm0.019$\\
        W1& 0.370 &$ -0.004\pm0.029$\\
        W2& 0.368 &$ -0.001\pm0.046$\\
        W3& 0.360 &$ +0.006\pm0.025$\\
        W4& 0.339 &$ +0.026\pm0.044$\\
        \noalign{\smallskip}
        \hline
        \noalign{\smallskip}
        \multicolumn{3}{@{}l}{Angular diameters (mas)} \\ 
        $\theta_1$ &$ 0.04721\pm  0.00036$ & $ -0.0000 \pm 0.0005 $\\
        $\theta_2$ &$ 0.02892\pm  0.00022$ & $ -0.0000 \pm 0.0003 $\\
        \noalign{\smallskip}
        \hline
    \end{tabular}
\end{center}
\end{table}

\begin{table}
\caption{Fundamental parameters of \cpd{} from the adopted light curve, radial velocity and \teff{} fits.
For comparison we also quote the values from \citet{2021MNRAS.500.4972R}. Quantities are given in nominal solar units \citep{2016AJ....152...41P}.}
\label{FundamentalParams}
\begin{center}
    \begin{tabular}{@{}lccc}
        \hline
        \noalign{\smallskip}
        \multicolumn{1}{@{}l}{Parameter} & Value & Value \\
        & (This work) & \citepalias{2021MNRAS.500.4972R} \\
        \noalign{\smallskip}
        \hline
        \noalign{\smallskip}
        $M_1$ (\si{\Msun})	&	$	1.3094	\pm 0.0051	$	&	$	1.311	\pm	0.035	$	\\
        $M_2$ (\si{\Msun})	&	$	1.0896	\pm	0.0034	$	&	$	1.093	\pm	0.029	$	\\
        $R_1$ (\si{\Rsun})	&	$   1.9288	\pm	0.0030	$   &	$	1.935	\pm	0.020	$	\\
        $R_2$ (\si{\Rsun})	&	$   1.1815	\pm	0.0037	$   &	$	1.181	\pm	0.014	$	\\
        $M_1+M_2$        	&	$	2.3990  \pm 0.0082  $	&			---			\\        
        $M_2/M_1$        	&	$	0.8321  \pm 0.0018  $	&			---			\\        
        $\log g_1$ (cm/s) 	&	$	3.9836	\pm	0.0012	$	&	$	3.982	\pm	0.006	$	\\
        $\log g_2$ (cm/s) 	&	$	4.3297	\pm	0.0026	$	&	$	4.332	\pm	0.008	$	\\
        $\rho_1$ ($\rho_{\odot}$) & $ 0.18207 \pm 0.00059 $ & --- \\
        $\rho_2$ ($\rho_{\odot}$) & $  0.6595 \pm  0.0059 $ & --- \\
        \teffpri{} (K)	&	$	6462	\pm	43	$	&	$	5980	\pm	205	$	\\
        \teffsec{} (K)	&	$	6331	\pm	43	$	&	$	5850	\pm	190	$ 	\\
        \teffsec{}/\teffpri{} & $ 0.9799 \pm 0.0023 $ & --- \\
        $\log L_1$ (\si{\Lsun})	&	$	0.766 \pm 0.011	$	&	$	0.635	\pm	0.059	$ \\  
        $\log L_2$ (\si{\Lsun})	&	$	0.305 \pm 0.012	$	&	$	0.168	\pm	0.060	$ \\  
        \noalign{\smallskip}
        \hline
    \end{tabular}
\end{center}
\end{table}

\begin{landscape}

\begin{table}
\caption{Fit results from different sets of input parameters. 
Values in parentheses are 1-$\sigma$ standard errors in the final digit(s) of the preceding value. N$_{\Delta}$ is the number of distortion coefficients included per star, $\Delta \lambda$ is the size of the integrating function wavelength bins in \AA, and $\log{\cal L}$ is the log-likelihood.
*N.B. these parameters have a non-Gaussian probability distribution. E(B$-$V) are given as 1-$\sigma$ upper limits.
}
\label{RunTable}
\begin{center}
    \begin{tabular}{@{}lrrrrrrrrrrrrl}
        \hline
        \noalign{\smallskip}
        \multicolumn{1}{@{}l}{Run} & 
        \multicolumn{1}{l}{T$_{\rm mod,1}$}& 
        \multicolumn{1}{l}{T$_{\rm mod,2}$}& 
        \multicolumn{1}{l}{[Fe/H]} &
        \multicolumn{1}{l}{[$\alpha$/Fe]} &
        \multicolumn{1}{@{}l}{$N_{\Delta}$} & 
        \multicolumn{1}{l}{$\Delta \lambda$}&
        \multicolumn{1}{l}{\teffpri{}}& 
        \multicolumn{1}{l}{\teffsec{}}&
        \multicolumn{1}{l}{E(B$-$V)*} &
        \multicolumn{1}{l}{$\sigma_{\rm ext,m}$*} &
        \multicolumn{1}{l}{$\sigma_{{\rm ext,}\ell}$*} &
        \multicolumn{1}{l}{$\log{\cal L}$} & 
        \multicolumn{1}{l}{Notes} \\
         & 
        \multicolumn{1}{c}{[K]}& 
        \multicolumn{1}{c}{[K]}& 
        \multicolumn{1}{c}{[dex]} &
        \multicolumn{1}{c}{[dex]} &
         & 
        \multicolumn{1}{c}{[\AA]}&
        \multicolumn{1}{c}{[K]}& 
        \multicolumn{1}{c}{[K]}&
        \multicolumn{1}{c}{[mag]} &
        \multicolumn{1}{c}{[mag]} &
        \multicolumn{1}{c}{[mag]} &
        & \\
        \noalign{\smallskip}
        \hline
        \noalign{\smallskip}
        A   &	6450	&	6300	&	0.0	&	0.0	&	3	&	50	&	$	6462	\pm	43	$	&	$	6331	\pm	43	$	&	0.0096	(72)	&	0.015	(16)	&	0.0029	(29)	&	89.0	&	Nominal	\\
        B	&	6650	&	6500	&	0.0	&	0.0	&	3	&	50	&	$	6463	\pm	42	$	&	$	6334	\pm	42	$	&	0.0094	(72)	&	0.011	(11)	&	0.0028	(28)	&	89.3	&	Different model SED parameters: \teff{}	\\
        C	&	6250	&	6100	&	0.0	&	0.0	&	3	&	50	&	$	6460	\pm	47	$	&	$	6326	\pm	46	$	&	0.0092	(70)	&	0.021	(23)	&	0.0031	(31)	&	86.0	&	"	\\
        D	&	6650	&	6100	&	0.0	&	0.0	&	3	&	50	&	$	6458	\pm	41	$	&	$	6321	\pm	40	$	&	0.0092	(70)	&	0.010	(12)	&	0.0036	(36)	&	88.3	&	"	\\
        E	&	6375	&	6375	&	0.0	&	0.0	&	3	&	50	&	$	6461	\pm	42	$	&	$	6333	\pm	43	$	&	0.0094	(70)	&	0.014	(15)	&	0.0031	(30)	&	88.7	&	"	\\
        F	&	6450	&	6300	&	-0.2	&	0.0	&	3	&	50	&	$	6476	\pm	41	$	&	$	6343	\pm	41	$	&	0.0098	(73)	&	0.010	(11)	&	0.0032	(33)	&	89.9	&	Different model SED parameters: [Fe/H]	\\
        G	&	6450	&	6300	&	0.2	&	0.0	&	3	&	50	&	$	6448	\pm	44	$	&	$	6316	\pm	44	$	&	0.0095	(71)	&	0.016	(19)	&	0.0029	(29)	&	87.4	&	"	\\
        H	&	6450	&	6300	&	0.0	&	0.0	&	3	&	50	&	$	6461	\pm	40	$	&	$	6335	\pm	41	$	&	0.0090	(69)	&	0.013	(12)	&	0.0028	(29)	&	89.5	&	BT-Settl-CIFIST model	\\
        I	&	6450	&	6300	&	0.0	&	0.0	&	3	&	50	&	$	6467	\pm	54	$	&	$	6312	\pm	101	$	&	0.0094	(72)	&	0.014	(14)	&	0.0150	(140)	&	66.9	&	Excluding PEST flux ratios in BVRI bands	\\
        J	&	6450	&	6300	&	0.0	&	0.0	&	3	&	50	&	$	6459	\pm	45	$	&	$	6333	\pm	59	$	&	0.0095	(71)	&	0.014	(15)	&	0.0051	(66)	&	64.2	&	Removing prior on near-IR flux ratios	\\
        K	&	6450	&	6300	&	0.0	&	0.0	&	3	&	50	&	$	6592	\pm	139	$	&	$	6456	\pm	133	$	&	0.0400	(320)	&	0.014	(15)	&	0.0029	(28)	&	89.1	&	Removing prior on E(B$-$V)	\\
        L	&	6450	&	6300	&	0.0	&	0.0	&	3	&	50	&	$	6462	\pm	43	$	&	$	6331	\pm	44	$	&	0.0095	(73)	&	0.015	(15)	&	0.0028	(27)	&	89.4	&	Removing prior on radius ratio $k$	\\
        M	&	6450	&	6300	&	0.0	&	0.0	&	0	&	50	&	$	6428	\pm	29	$	&	$	6298	\pm	30	$	&	0.0039	(31)	&	0.013	(11)	&	0.0026	(24)	&	84.2	&	SED fitting (no distortion)	\\
        N	&	6450	&	6300	&	0.0	&	0.0	&	6	&	50	&	$	6463	\pm	45	$	&	$	6330	\pm	44	$	&	0.0090	(68)	&	0.018	(16)	&	0.0041	(45)	&	89.7	&	Varying N$_\Delta$	\\
        O	&	6450	&	6300	&	0.0	&	0.0	&	9	&	50	&	$	6472	\pm	49	$	&	$	6329	\pm	53	$	&	0.0094	(72)	&	0.020	(18)	&	0.0080	(100)	&	88.7	&	"	\\
        P	&	6450	&	6300	&	0.0	&	0.0	&	3	&	20	&	$	6465	\pm	43	$	&	$	6334	\pm	43	$	&	0.0101	(75)	&	0.013	(14)	&	0.0028	(27)	&	89.2	&	Different wavelength binning	\\
        Q	&	6450	&	6300	&	0.0	&	0.0	&	3	&	80	&	$	6463	\pm	42	$	&	$	6331	\pm	43	$	&	0.0096	(72)	&	0.016	(16)	&	0.0031	(34)	&	88.9	&	"	\\
        \noalign{\smallskip}
        \hline
    \end{tabular}
\end{center}
\end{table}
\end{landscape}

\section{Discussion}\label{sec:Discussion}

\subsection{Impact of third light}
 We are not confident that the value of $\ell_3\approx 1$\,per~cent found from the analysis of the \tess{} light curve represents a genuine detection of flux from a third body in the system. We suspect that this value is more likely due to systematic errors in the values of the background flux level in the \tess{} images and/or an underestimated contamination of the photometric aperture by other stars in the image. Nevertheless, we should consider the possibility that this flux is due to a low-mass tertiary star in the system and estimate the impact on our effective temperature measurements. The absolute magnitude of \cpd\ in the \gaia{} RP band is  M$_{\rm RP} = 2.071$. The \gaia{} RP band is similar to the \tess{} bandpass so we can estimate that the absolute magnitude of the putative third body is M$_{\rm RP} \approx 7.1$. Assuming that the third body is a main-sequence star, this corresponds to a K9V star with a luminosity of 0.066\,L$_{\odot}$, i.e. 0.8\,per~cent of the total luminosity.\footnote{\url{http://www.pas.rochester.edu/~emamajek/EEM_dwarf_UBVIJHK_colors_Teff.txt}} Assuming that the extra luminosity is assigned equally between the two stars by our method if it is not accounted for, the effective temperatures we have measured will be over-estimated by 9\,K for the primary star and 25\,K for the secondary star. This is a negligible effect when compared to the standard errors on the values.  
 
\subsection{Comparison to stellar evolution models}
 We have compared the properties of \cpd\ to stellar evolution tracks computed with the Garching Stellar Evolution Code \citep[GARSTEC][]{2008Ap&SS.316...99W}.  The microphysics used in these models is described in \cite{2013MNRAS.429.3645S} and \cite{2008Ap&SS.316...99W}, but we provide a very brief summary here. The convection is described by the standard mixing length theory of \cite{1990sse..book.....K}, where the solar mixing length is $\alpha_{\rm ml,\odot} = 1.801$ using the \cite{1993oee..conf...15G} solar composition. The models include convective mixing and convective overshooting described in terms of diffusive processes. Due to the effects of diffusion, the initial solar composition is found to be ${\rm [Fe/H]}_{\rm i}= +0.06$.
 
 Several grids of models were computed varying either the initial helium abundances or the assumed mixing length. Each grid covers stellar masses in the range 0.7-2.0\,M$_{\sun}$ in steps of 0.02\,M$_{\sun}$ and ages from the zero-age main sequence up to $\tau=$17.5\,Gyr. For each grid of models we used a Markov-chain Monte Carlo  method to sample the posterior probability distribution of the model parameters $P(\tau$, $M_1$, $M_2$, ${\rm [Fe/H]}_{\rm i}|D)$, where the data $D$ are the fundamental parameters of the stars given in Table~\ref{FundamentalParams}. Further details of the stellar evolution models and MCMC methods used are provided in \citet{2018A&A...615A.135K}.
 
\begin{figure}
    \centering
    \includegraphics[width=0.49\textwidth]{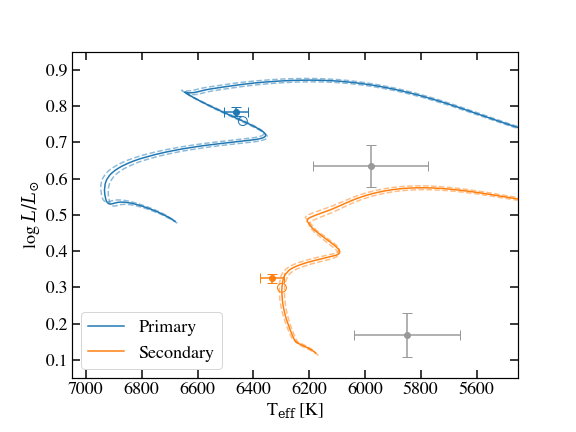}
    \caption{\cpd\ in the Hertzsprung-Russell diagram compared to GARSTEC stellar evolution models. Error bars in grey are the parameters from \citet{2021MNRAS.500.4972R}. GARSTEC stellar evolution tracks are shown for masses $M_1  = 1.308\pm0.005\,M_{\odot}$ and $M_2  = 1.085\pm0.003\,M_{\odot}$, initial metallicity [Fe/H]$_{\rm i} = -0.05$, initial helium abundance $Y_{\rm i} = Y_{{\rm i},\odot}+0.03$ and a mixing length $\alpha_{\rm MLT}=1.78$. The open circles on the evolutionary tracks correspond to the best-fit age of 2.83\,Gyr. } 
    \label{fig:modvobs2}
\end{figure}

 We found the best fit to the observed parameters of \cpd\ for the grid of models with an initial helium abundance 0.03 dex higher than our assumed solar initial helium abundance and a mixing length $\alpha_{\rm ml}=1.78$ at an age $\tau = 2.83$ Gyr. The best-fit stellar evolution tracks are shown in the Hertzsprung-Russell diagram in Fig.~\ref{fig:modvobs2}. For the best fit we obtain $\chi^2=8.2$ for 7 observed quantities and 4 free parameters, i.e. 3 degrees of freedom. Models of solar initial helium abundance give a significantly worse fits to the observations ($\chi^2>50$). However, we note that the primary star is near the end of its main-sequence evolution where the evolution models are sensitive to assumptions about the nature and extent of mixing processes near the core, so there may be other models that fit the observations equally well assuming an initial helium abundance closer to the solar value. A full exploration of the parameter space for stellar models is beyond the scope of this study, but is certainly worthwhile given the high precision and accuracy of the fundamental parameters for these stars that are now available.
 
This analysis shows that the parameters we have obtained are consistent with stellar evolution models for a normal pair of main-sequence stars slightly younger than the Sun with the primary close to the end of its  main-sequence lifetime. It also suggests that the high-precision and accuracy of the stars' fundamental parameters we have derived, particularly the effective temperatures, provide useful constraints on model parameters such as the mixing length and initial helium abundance. A full exploration of the model parameter space is needed to quantify the precision of the model parameters that can be derived from such a comparison and the correlations between them, but is beyond the scope of this study.
  
\subsection{Requirements for precise and accurate effective temperature estimates}\label{sec:teff-discussion}

\subsubsection{Handling of `incorrect' choice of model with distortion}
To verify the ability of our method to handle a range of reasonable starting values without the result changing significantly, we ran the method with different starting temperatures (runs B--E) and metallicities (runs F--G) for the BT-Settl model SEDs than those used to derive the values given in Table~\ref{FundamentalParams}. Increasing the model temperatures for both stars by 200\,K (run B) gives a near-indistinguishable result, but decreasing both model temperatures by the same amount (run C) slightly increases the errors in derived temperatures and increases the amount of additional noise given to the observed magnitudes. 
Increasing and decreasing the ratio between the model temperatures for the two stars (runs D and E respectively) again shows no significant difference to the adopted values, although in run D we can see that the code tries to compensate for the increased temperature ratio by boosting the additional noise for the flux ratios and reducing the estimate for interstellar reddening, which in turn results in a derived secondary star temperature 10\,K cooler than our adopted value. Varying the ratio of the input temperatures is a useful test to check whether the distortion is behaving correctly, and given the agreement of runs D and E with our adopted values from run A, we do see that the method of distortion works. 
Since our estimate for the metallicity of \cpd{} has an uncertainty of $\pm0.2$\,dex, we tested whether an input [Fe/H] value at either end of this range (runs F and G) would have a significant impact on the derived effective temperatures. We see a slight increase of 10-15\,K in the derived \teff{} for both components when $\rm{[Fe/H]}=-0.2$, and a similar decrease for both components when $\rm{[Fe/H]}=0.2$. At a lower metallicity, the amount of line-blanketing in the near-ultraviolet region increases in the model SEDs. To compensate for this, the distortion functions must boost the amount of flux in the ultraviolet, and hence increasing the derived effective temperatures. This highlights how important it is to not only have a reliable estimate for $\rm{[Fe/H]}$, but also photometry in the near-ultraviolet region to provide a reasonable constraint on the fit. 
Finally, we also tested whether using a different set of model SEDs (BT-Settl-CIFIST rather than BT-Settl, run H) with different abundances would have an impact on the derived effective temperatures. We see no significant change in the derived temperatures or goodness-of-fit metrics, which suggests that there is no significant dependence on the choice of model SED.
From these tests, we have shown that the temperature method is robust to different input model SEDs generated using a reasonable range of temperatures and metallicities. The distortion handles ``incorrect'' temperature ratios that a simple SED fit would be unable to, and these results show that the distortion breaks any strong dependence on model choice, i.e. the effective temperatures we have derived are not strongly dependent on the details of how the analysis has been done.

\subsubsection{Value of multi-band light curves}

\cpd{}, like many eclipsing binary systems in the literature, does not have light curves in passbands beyond the broad-band visible or NIR of large photometric surveys such as ASAS, WASP, \textit{Kepler} and \tess{}. While it is possible to obtain an estimate of \teff{} with only one binary flux ratio, the distortion functions are less well constrained throughout the optical range and the results are therefore less precise. Obtaining light curves of \cpd{} during the total secondary eclipse made it possible to add an additional four constraints on the binary flux ratio throughout the visible range, and tightened up the uncertainties on \teff{}: when running the \teff{} fit with the same parameters as the adopted run but excluding PEST flux ratios (Run I), the uncertainties on \teff{} increase from $6462\pm43$\,K to $6467\pm54$\,K for the primary and $6331\pm43$\,K to $6311\pm101$\,K for the secondary. Given the importance of obtaining robust and precise direct measurements for the \teff{}, this comparative test of the method with and without the PEST flux ratios highlights the importance of including as much multi-band photometric data as possible.

\subsubsection{Priors on parameters}

We tested the impact of each of the additional priors we placed on model parameters for the temperature fit: near infrared flux ratios (run J), interstellar reddening (run K), and ratio of the fractional stellar radii (run L). 
As with AI Phoenicis \citepalias{2020MNRAS.497.2899M}, placing a prior on the relative flux of the two component stars in the near-infrared was a useful addition to the code that prevented any potential distributions of flux about the wavelength range that was non-physical. Removing this prior in run J gives no significant change in the value of \teff{} for either component, but the effect can be seen in the increased amount of additional noise used to fit the observed flux ratios, and slightly increased uncertainty in the effective temperature for the secondary component ($\sim16$\,K). These results are encouraging and support the use of this prior on the near-infrared flux ratio for systems like \cpd{} as part of the standard procedure for the temperature method in future work.

 To derive precise and accurate effective temperatures with our method, it is essential to have a reliable, direct estimate for E(B$-$V). This is demonstrated in run K, for which we relaxed the prior we placed on E(B$-$V) from the Na D I line equivalent width. Both the values and uncertainties for the derived temperatures of both components in \cpd{} have significantly increased from the adopted run A to this run, K. The code climbs to a much higher yet more uncertain value for E(B$-$V) without the constraint from the prior, which in turn makes it necessary to increase the derived temperatures.
In contrast, the prior on the radius ratio has very little impact on the results or quality of fit for \cpd{} (run L).

\subsubsection{Number of distortion coefficients}

The measurements of the flux ratios are least affected by systematic errors because they are directly related to the eclipse depth, and the SED for both stars are very similar, so errors in zero-points and instrumental response functions affect both stars equally. We can therefore inspect how well the code treats the flux ratio measurements and use this to select the optimal number of distortion coefficients; i.e. the fit with the lowest $\sigma_{\rm ext,\ell}$. An inflated value for $\sigma_{\rm ext,\ell}$ can be due to over-fitting -- the wavelength space is poorly sampled due to very few measurements of flux ratio -- or due to SED fitting, where there is little else the code can vary to optimise the overall fit. 
The results for the temperature fit, using the method with no distortion to illustrate a standard SED fitting procedure, is shown as run M. While the errors on derived \teff{} for both components are lower than all other runs in Table \ref{RunTable}, we argue that these are underestimates of the true errors. The additional noise required to fit the observed flux ratios is double that which was required for the adopted run A, and the log-likelihood is much lower, which indicates a worse fit.
We thoroughly tested which number of distortion coefficients was optimal for \cpd{}. The results for $N_{\Delta}=3$ (adopted values, run A), 6 (run N) and 9 (run O) are shown in Table \ref{RunTable}. The uncertainties on derived temperatures increase with an increasing number of coefficients, to the point of potential over-fitting with run O. We settled on 3 sets of coefficients for \cpd{}, balancing the need for distortion whilst avoiding over-fitting due to the sparse photometric data available. 
For stars with more photometric data throughout the log-wavelength space, such as AI Phoenicis, it is more reasonable to use a larger number of coefficients to ensure the models can be distorted on a small enough scale to accurately fit the observational data.
With spectrophotometry across the optical range from the upcoming \gaia{} DR3 and beyond, it may be possible to employ a greater number of distortion coefficients for the \teff{} fitting method.

\subsubsection{Effects of model SED binning}
We tested whether the model SED being binned into smaller or larger wavelength bins would have an effect on the results of the \teff{} fit. Run P has the most fine grid at 20\,\AA, Run Q has the largest bin size of 80\,\AA~and these can both be compared to the adopted Run A with a bin size of 50\,\AA. The output of the \teff{} fit in all three scenarios is largely the same, with the log-likelihood growing slightly with smaller wavelength bins. The main difference between these three fits is the run time. For 2000 steps with 256 walkers on a standard desktop computer with 4 Intel Core i5-7500 CPUs at 3.40\,GHz, the most fine wavelength grid of 20\,\AA~(run P) takes 2 hours 13 minutes, the adopted run with a 50\,\AA~grid takes 1 hour 16 minutes, and the most coarse grid we tested (run Q at 80\,\AA) takes 1 hour 04 minutes. We therefore conclude that the choice in wavelength bin size is ultimately a balance between minimising run time and maximising the quality of the fit. For the adopted and other runs we settled for 50\,\AA~as a compromise between these factors.

\subsection{Effective temperature estimates from disentangled spectra}
When compared to the spectroscopic temperature estimates derived in \citetalias{2021MNRAS.500.4972R}, the fundamental effective temperatures measured in this work are significantly hotter for both components (see Table \ref{FundamentalParams}). Despite the difference in \teff{} estimates between the two analyses, we can still obtain a synthetic spectrum fit of a similar quality with the higher temperatures (see Figure \ref{fig:synthetic_spectrum} versus Figure 1 in \citetalias{2021MNRAS.500.4972R}). This highlights the difficulties of obtaining reliable \teff{} estimates from low to moderate signal-to-noise spectra, and stresses the importance of not only developing a catalog of stars with accurate fundamental effective temperature measurements, but of independently checking spectroscopic effective temperature estimates with other methods where possible.

\section{Conclusions}

We have measured precise and accurate masses, radii and effective temperatures for both stars in the detached eclipsing binary \cpd{} using a wealth of new \tess{} light curves, radial velocities measured by \citetalias{2021MNRAS.500.4972R}, observations of the total secondary eclipse in BVRI bands, photometry from the ultraviolet to near-infrared, and parallax from \gaia{} EDR3. Using the method first described in \citetalias{2020MNRAS.497.2899M}, we have significantly improved the measured values for effective temperature. We find that the stars in \cpd{} are slightly younger than the Sun, with the primary F5\,V component appearing to be close to the end of its main-sequence lifetime.
\cpd{} is a detached, well-behaved and isolated system which makes it ideal for testing calibrating data-driven stellar parameter pipelines from spectroscopic surveys, along with testing stellar evolution models. There are many more moderately-bright systems like \cpd{} being discovered by large scale surveys such as \tess{} which are suitable candidate benchmark stars for future work building on this method for deriving direct, accurate effective temperatures for stars in detached eclipsing binaries.

\section*{ACKNOWLEDGEMENTS}

NM and PM are supported by the UK Science and Technology Facilities Council (STFC) grant numbers ST/M001040/1 and ST/S505444/1.
TGT (PEST) wishes to thank Craig Bowers for the loan of the ST-8XME camera used in this research.
This research has made use of the SIMBAD database, operated at CDS, Strasbourg, France \citep{2000A&AS..143....9W}. 
Based on observations collected at the European Southern Observatory under ESO programmes 089.D-0097(A) and 090.D-0061(A).
This research made use of Lightkurve, a Python package for \textit{Kepler} and \textit{TESS} data analysis \citep{2018ascl.soft12013L}.
This research made use of Astropy,\footnote{http://www.astropy.org} a community-developed core Python package for Astronomy \citep{astropy:2013, astropy:2018}.
We would like to thank the anonymous referee for their constructive and timely comments on the manuscript.

\section*{DATA AVAILABILITY}

The data used in this article are available from the following sources:
Table \ref{tab:ground-based-photometry} of this work (WASP and PEST photometry);
Mikulski Archive for Space Telescopes -- \url{https://archive.stsci.edu/} (TESS);
ESO Archive Science Portal -- \url{http://archive.eso.org/scienceportal/home} (FEROS);
Table A1 of \citet{2021MNRAS.500.4972R} (radial velocities from FEROS, CHIRON, CORALIE spectra).



\bibliographystyle{mnras}
\bibliography{cpd-54810} 




\appendix

\section{Alternative \tess{} light curve fits}

\subsection{Analysis with \jktebop{}}\label{jktebop}
For this analysis, we re-analysed the light curve of \cpd{} using all suitable \tess{} data that is currently available, including the newer 10-minute cadence observations.
The two sets of observations (2-minute and 10-minute cadences) were split further into 5 and 4 sections respectively, containing at least one primary and one secondary eclipse, which were all analysed separately.
We performed light curve fits for each section with \jktebop{}\footnote{Version 40. The code is available at \url{https://www.astro.keele.ac.uk/jkt/codes/jktebop.html}} \citep{2013A&A...557A.119S}, which uses Levenberg-Marquardt minimisation \citep{1992nrfa.book.....P} to find the optimal solution for the \textsc{ebop} light curve model \citep{1981AJ.....86..102P, 1981ASIC...69..111E}.
We used the quadratic limb darkening law for both components of \cpd{}.
The free parameters in each fit were: the surface brightness ratio in the \tess{} band $J = S_{\rm T,2}/S_{\rm T,1}$, sum of the fractional radii $r_{\rm sum} = r_1+r_2 = R_1/a + R_2/a$, ratio of the fractional radii $k = r_2/r_1$, the quadratic limb darkening coefficients (where the coefficient for the secondary star was set as equal to those for the primary), orbital inclination $i$, $e\cos{\omega}$, $e\sin{\omega}$, third light $\ell_3$, and the light scale factor. 
The values of orbital period $P$ and time of primary minimum $T_0$ were fixed at the best values from the calculation of the linear ephemeris in Section \ref{sec:ephem}.
The mean and standard error for each free parameter was calculated from all nine sections of \tess{} observations and taken as the adopted solution. 
This approach is justified by \citet{2020MNRAS.498..332M}, in which the authors demonstrate that the MC and RP errorbars in \jktebop{} are reliable, and in \citet{2021Obs...141..190S}, where the author shows that these MC and RP errorbars agreed with those obtained from fitting the data in subsets.
The best values for each parameter fitted by \jktebop{} are shown in Table \ref{LC-RV-results}.

\begin{table}
\caption{Orbital elements of \cpd{} from the \jktebop{} fits to the \tess{} light curves and \textsc{ellc} radial velocity fits. The quadratic limb darkening coefficients $c_1$, $c_2$ are the same for both stars.}
\label{LC-RV-results}
\begin{center}
    \begin{tabular}{@{}lcc} 
    \hline
    \noalign{\smallskip}
    \multicolumn{1}{@{}l}{Parameter} & Light curves & Radial velocities \\
    \noalign{\smallskip}
    \hline
    \noalign{\smallskip}
        $J$ & 0.9372 $\pm$ 0.0020 & -- \\
        $r_{\rm sum}$		&	0.06272	$\pm$	0.00003	&		--		\\
        $k$		&	0.6110	$\pm$	0.0013	&		--		\\
        $c_{1}$		&	0.32	$\pm$	0.03	&		--		\\
        $c_{2}$		&	0.09	$\pm$	0.05	&		--		\\
        $i$	($^{\circ}$)	&	89.72	$\pm$	0.021	&		--		\\
        $\ell_3$ & 0.002 $\pm$ 0.004& -- \\
        $e$		&	0.3686	$\pm$	0.0001	&	0.3683	$\pm$	0.0006	\\
        $\omega$	($^{\circ}$)	&	327.02	$\pm$	0.03	&	327.18	$\pm$	0.17	\\
        K$_1$	(km\,s$^{-1}$)	&		--		&	46.93	$\pm$	0.06	\\
        K$_2$	(km\,s$^{-1}$)	&		--		&	56.40	$\pm$	0.10	\\
        $\gamma_1$	(km\,s$^{-1}$)	&		--		&	0.38	$\pm$	0.05	\\
        $\gamma_2$	(km\,s$^{-1}$)	&		--		&	0.56	$\pm$	0.07	\\
    \noalign{\smallskip}
    \hline
    \end{tabular}
\end{center}
\end{table}

We performed a new fit of the radial velocities extracted by \citetalias{2021MNRAS.500.4972R} using the radial velocity model in \textsc{ellc} \citep{2016A&A...591A.111M}. We allowed the following parameters to be free: $K_1$, $K_2$, $\gamma_1$, $\gamma_2$, $T_0$, period $P$, $\sqrt{e}\cos{\omega}$, $\sqrt{e}\sin{\omega}$ and the excess noise in the radial velocities $\sigma_{\rm rv}$.  
We placed Gaussian priors on $T_0$ and $P$ from the ephemeris derived in section \ref{sec:ephem}. The posterior probability distribution of the model parameters was sampled using the {\sc emcee} implementation of the affine-invariant ensemble sampler for Markov chain Monte Carlo \citep{2013PASP..125..306F}, using 512 walkers over a chain of 600 steps and burn-in of 400 steps. The model parameters derived are given in Table \ref{LC-RV-results}.
We also did a least-squares fit to the radial velocity data including priors on $e$ and $\omega$ from the analysis of the light curves. The results were almost identical to those presented in Table~\ref{LC-RV-results} so we do not report them here. This insensitivity to the exact choice of $e$ and $\omega$ is because the radial velocity curves for both stars are well sampled around their minima and maxima.

\begin{figure}
    \centering
    \includegraphics[width=0.475\textwidth]{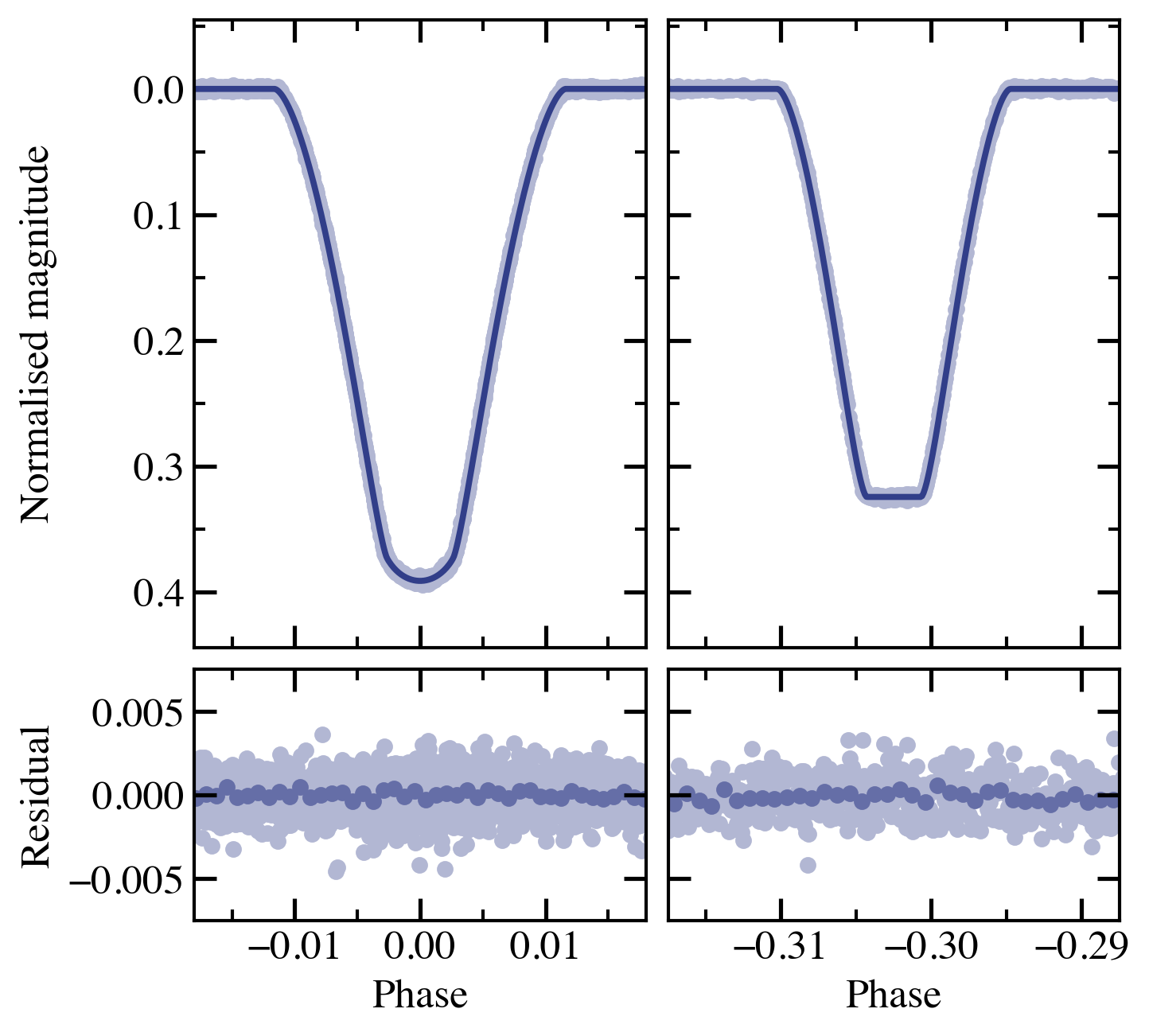}
    \caption{Primary and secondary eclipses of \cpd{} in Sectors 6 and 7, as observed by \tess{} in the 2-minute cadence. The best model from \jktebop{} for this section of the 2-minute data se and the residual of the fit are shown alongside the raw photometry data.}
    \label{fig:tess}
\end{figure}

\subsection{Analysis with the WD code}
\label{wd}
For the analysis we used 2-minute cadence data from 5th, 6th and 10th \tess{} sectors. The data contain three primary and three secondary eclipses. We retained all points within eclipses and just around them and each 40th point in out-of-eclipse parts of the light curve. The light curve was detrended from a long-term small light variations. Its out-of-eclipse parts are practically flat. In total 3834 data points were used. For radial velocities we adopted RVs published by \citetalias{2021MNRAS.500.4972R}. We used all their RVs with an exception of one measurement taken at HJD 2456400.485 (at the orbital phase 0.20), which deviates significantly from the model. Our naming of components is reversed to that used by \citetalias{2021MNRAS.500.4972R} as we call the primary a more massive, larger and brighter component which is eclipsed during a deeper minimum.

Simultaneous analysis of light and radial velocity curves was performed with the Wilson-Devinney (WD) code version 2015 \citep{1971ApJ...166..605W, 1979ApJ...234.1054W, 1990ApJ...356..613W, 2007ApJ...661.1129V, 2014ApJ...780..151W} equipped with the Python GUI written by \citet{2020CoSka..50..535G}. The latest WD version allows for a direct modeling of photometry obtained in the \tess{} filter (number 95 in the WD) and a high numerical precision. The orbital period was set to the value derived from analysis of minima times (see Section \ref{sec:ephem}), the surface grid parameters (the numerical precision) was set to \verb"N1=N2=80". The limb darkening coefficients were fixed to values from updated tables originally published by \citet{1993AJ....106.2096V} according to actual values of surface gravity and temperature at fixed metallicity of [Fe/H]=0 (see Section \ref{sec:extinction}). During analysis both the logarithmic \citep{1970AJ.....75..175K} and square root \citep{1992A&A...259..227D} limb-darkening (LD hereafter in this section) laws were tested. The albedo parameters were set to 0.5 and the gravity brightening parameters were set to 0.32 for both components as their atmospheres are expected to be fully convective. The atmosphere approximation was used \verb"IFAT1=IFAT2=1" and the radial velocity corrections were applied \verb"ICOR1=ICOR2=1". While modeling the following parameters were allowed to vary: the orbital phase shift, the luminosity of the primary $L_1$, the eccentricity $e$, the longitude of periastron $\omega$, the mass ratio $q$, the semi-major axis $a$, the orbital inclination $i$, the dimensionless Roche potentials $\Omega$, the temperature of the secondary $T_2$, the systemic velocities $\gamma$ and also the third light $\ell_3$. After few iteration the phase shift was fixed at 0.1136 and then the epoch of the primary minimum T$_0$ was adjusted during later analysis.

\begin{table}
\caption{Photometric and orbital parameters of \cpd{} from the WD fits to the \tess{} light curves and the radial velocities.}
\label{WD-results}
\begin{center}
    \begin{tabular}{@{}lcc}
    \hline
    \noalign{\smallskip}
    \multicolumn{1}{@{}l}{Parameter} & Value & Comments \\
    \noalign{\smallskip}
    \hline
    \noalign{\smallskip}
        Period	(d)	&	26.13132764		&	fixed	\\
        T$_0$	(d)$^1$	&	8470.10157	$\pm$ 0.00024 & 	\\
        $\Omega_1$		&	27.054	$\pm$	0.043	&				\\
        $\Omega_2$		&	36.397	$\pm$	0.145	&			\\
        $T2$ (K) & 6359 $\pm$ 3 &  \\
        $i$	($^{\circ}$)	&	89.825	$\pm$	0.030	&			\\
        $e$		&	0.3691	$\pm$	0.0001	&		\\
        $\omega$	($^{\circ}$)	&	326.86	$\pm$	0.03	&	\\
        $a$ (\si{\Rsun}) & 49.718 $\pm$ 0.090  & \\
        $q$ & 0.8317 $\pm$ 0.0028 & \\
        $\gamma_1$	(km\,s$^{-1}$)	&	0.40	$\pm$	0.04 &	\\
        $\gamma_2$	(km\,s$^{-1}$)	&	0.39	$\pm$	0.05 &	\\
        $\ell_3$ (TESS) & 0.0130 $\pm$ 0.0025 & \\
        \multicolumn{3}{c}{Derived parameters}\\
        K$_1$	(km\,s$^{-1}$)	&	47.03 $\pm$ 0.09		&		\\
        K$_2$	(km\,s$^{-1}$)	&	56.54 $\pm$ 0.14	&	\\
        $r_1$		&	0.03886	$\pm$	0.00007	&		\\
        $r_2$		&	0.02395	$\pm$	0.00010	&			\\
        $r_{\rm sum}$		&	0.06281	$\pm$	0.00005	&			\\
        $k$		&	0.6164	$\pm$	0.0033	&		\\
        $L_2/L_1$ (TESS) & 0.3596 $\pm$ 0.0042 & direct\\
        $L_2/L_1$ (V) & 0.3519 & extrapolated \\
        $L_2/L_1$ (K$_{\rm 2MASS}$) & 0.3735 & extrapolated\\
    \noalign{\smallskip}
    \hline
    \end{tabular}
\begin{tablenotes}
$^1$T$_0$ is measured in BJD$-2450000$.
\end{tablenotes}
\end{center}
\end{table}

\begin{figure}
    \includegraphics[width=0.5\textwidth]{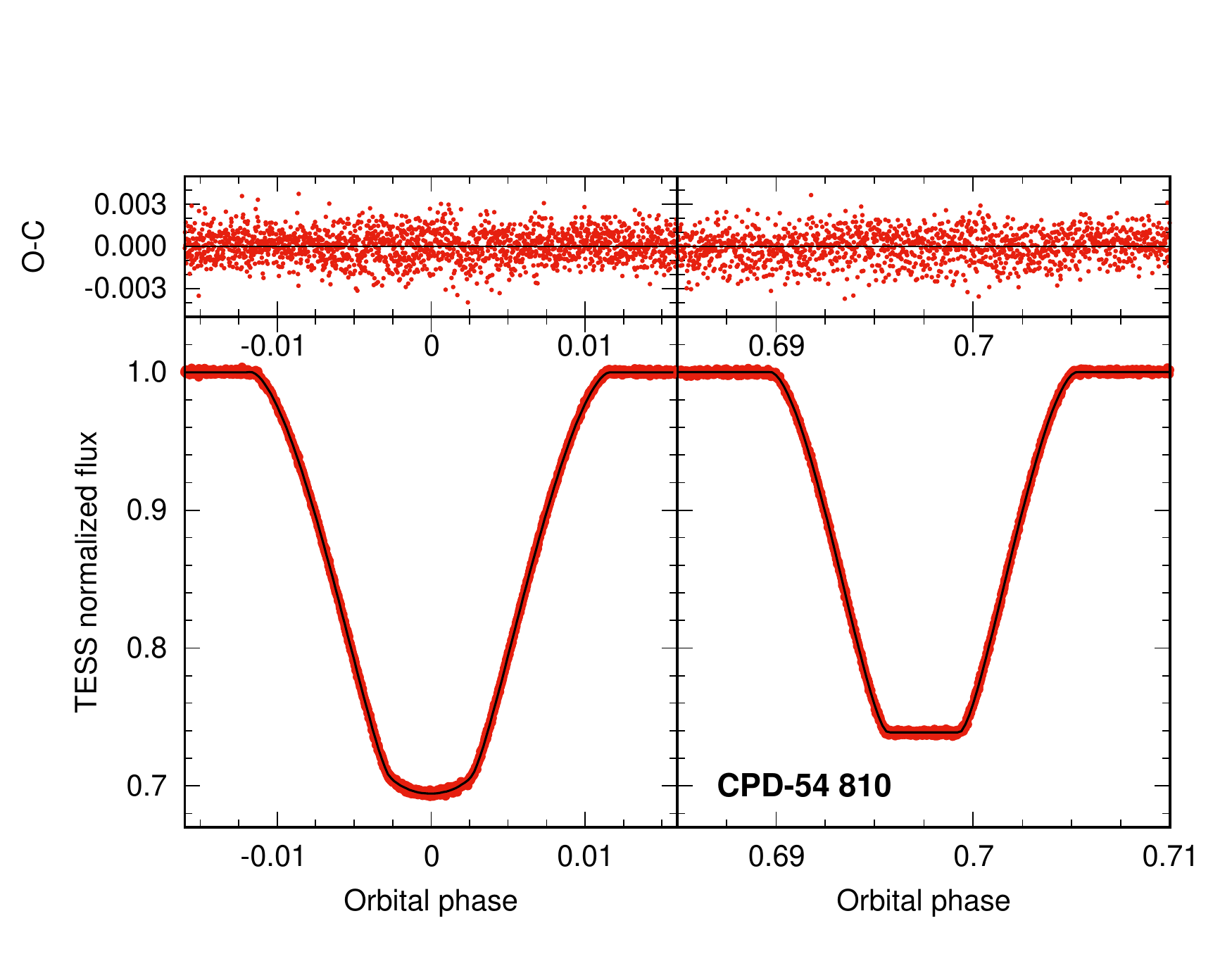}
    \caption{The WD fit to \tess{} light curve from 5th, 6th and 10th sectors.} 
    \label{fig:WD}
\end{figure}

We started the analysis by adopting as an input the model parameters reported by \citetalias{2021MNRAS.500.4972R}. In the beginning the logarithmic LD law was used and no third light was assumed. The resulted solution produced small but systematic residuals in both eclipses. A use of the square root LD law improved the {\it rms} of \tess{} light curve solution but the systematics were still present. Finally, adjusting $\ell_3$ produced acceptable fit to both eclipses without any noticeable systematics in residuals. The detected third light in \tess{} is small, at 1.3\%. It is not clear if that detection is real or results from using the LD law which is not fully adequate in case of both components. 

Mean errors on the radial velocity determination are 145 m\,s$^{-1}$ and 330 m\,s$^{-1}$ \citepalias{2021MNRAS.500.4972R}, while the obtained solution for radial velocities has the {\it rms} 175 m\,s$^{-1}$ and 273 m\,s$^{-1}$ for the primary and the secondary, respectively. The primary shows the slightly larger {\it rms} then expected which might be attributed to an influence of a putative third body in the system. The systemic velocities of both components are very similar and they differ no more than 40 m\,s$^{-1}$.  

In Table~\ref{WD-results} we summarised parameters of the best fit model. Errors quoted are formal errors returned by the Differential Corrections procedure but multiplied by a factor of three. For some parameters like fractional radii $r_1$ and $r_2$ they are much larger than those reported from the analysis with \jktebop{} or {\tt ellc}. The difference comes from very conservative errors adopted in case of the WD analysis but also from correlations between model parameters especially between the sum of the radii $r_{\rm sum}$, the orbital inclination $i$ and $\ell_3$. 

\subsection{Analysis with {\tt ellc}}
 We used the binary star model  {\tt ellc} \citep{2016A&A...591A.111M} to analyse every sector of \tess{} data containing both a complete primary and complete secondary eclipse, viz. sectors 4 to 7, 10 and 13 at 2-minute cadence, and sectors 27, 31 and 32 and 10-minute cadence. 
 Only data within one eclipse width of the phase of mid-eclipse were included in the analysis. 
 The data for each eclipse were divided by a straight line fit to the data either side of eclipse to remove instrumental trends prior to analysis. 
 We used the power-2 limb darkening law for both stars assuming that same values of $h_1$ and $h_2$ \citep[as defined in][]{2018A&A...616A..39M} for both stars. 
 We used the ``fine'' grid for the numerical integration of the fluxes through the eclipses so that the numerical noise is well below 20\,ppm at all phases. 
 The mutual gravitational distortion of the stars has a negligible impact on the light curve so we assumed spherical stars for the calculation of the model light curves. 
 For the 10-minute cadence data we used numerical integration to account for the finite integration time. 
 The orbital period was fixed at the value $P=26.131328$\,d. 
 The free parameters in the fit were: $J = S_{\rm T,2}/S_{\rm T,1}$, $r_{\rm sum} = r_1+r_2$, $k = r_2/r_1$, $i$, $f_c=\sqrt{e}\,\cos{\omega}$, $f_s = \sqrt{e}\,\sin{\omega}$, $T_0$, $h_1$, $h_2$, third light $\ell_3$, and a scaling factor.
 We used \textsc{emcee} \citep{2013PASP..125..306F} to find the mean and standard error of these parameters in the posterior probability distribution (PPD) assuming Gaussian white noise for the data. 
 The standard deviation per point was included as a hyperparameter when sampling the PPD. 
 Broad uniform  priors were applied to all parameters. For  $\ell_3$, negative values were permitted to allow for systematic errors in background subtraction and/or star spots. The PPD was sampled using 100 walkers running for 500 steps after discarding a ``burn-in'' phase of 1500 steps. 
 Convergence of the chains was verified by visual inspections of parameter values as a function of step number.
 No trends or excess noise during the eclipses was apparent from a visual inspection of the residuals from the best fit for all sectors. 
 The weighted mean and standard error of the weighted mean for the main parameters of interest are given in Table~\ref{compare-results}. 
 The mean values of the limb-darkening parameters are  $h_1 = 0.820 \pm 0.001$, $h_2 = 0.44\pm 0.02$. 
 These values agree well with the values expected based on STAGGER-grid 3-D atmosphere models \citep{2015A&A...573A..90M} given the effective temperature, surface gravity and metalicity of the two stars \citep[Star 1: $h_1 = 0.826$, $h_2 = 0.409$; Star 2: $h_1 = 0.813$, $h_2 = 0.429$,][]{2016A&A...591A.111M}.


\bsp	
\label{lastpage}
\end{document}